\documentclass[aps,prd,showpacs,onecolumn,preprintnumbers,notitlepage,
nofootinbib,amsfont,amssymb,10pt,singlespacing] {revtex4-1}
\usepackage{amsmath,bm}
\usepackage{times}
\usepackage{braket}
\usepackage{color,graphicx}
\usepackage{slashed}
\usepackage{hyperref}

\newcommand{\beq}{\begin{eqnarray}}
\newcommand{\eeq}{\end{eqnarray}}
\newcommand{\non}{\nonumber\\ }
\newcommand{\ov}{\overline}

\newcommand{\acp}{ A_{\rm CP}}
\newcommand{\cp}{{\it CP}}

\newcommand{\psl}{ P \hspace{-2.8truemm}/ }
\newcommand{\nsl}{ n \hspace{-2.2truemm}/ }
\newcommand{\vsl}{ v \hspace{-2.2truemm}/ }
\newcommand{\epsl}{\epsilon \hspace{-1.8truemm}/\,  }

\def\lsim{ {\ \lower-1.2pt\vbox{\hbox{\rlap{$<$}\lower6pt\vbox{\hbox{$\sim$}
}}}\ } }
\def\gsim{ {\ \lower-1.2pt\vbox{\hbox{\rlap{$>$}\lower6pt\vbox{\hbox{$\sim$}
}}}\ } }

\def \jhep{ J. High Energy Phys.  }

\definecolor{Red}{rgb}{1.,0.,0.}

\definecolor{Blue}{rgb}{0.,0.,1.}
\newcommand{\Blue}[1]{{\color{Blue}{#1}}}

\definecolor{nicered}{rgb}{0.7,0.1,0.1}
\definecolor{nicegreen}{rgb}{0.1,0.5,0.1}

\hypersetup{colorlinks,citecolor=nicegreen,linkcolor=nicered}

\begin{document}

\title{ Charmless hadronic $B \to (f_1(1285),f_1(1420)) P$ decays
in the perturbative QCD approach}

\author{Xin~Liu$^{1}$\footnote{Electronic address: {\Blue{liuxin.physics@gmail.com}}}}

\author{Zhen-Jun~Xiao$^{2}$\footnote{Electronic address: {\Blue{xiaozhenjun@njnu.edu.cn}}}}

\author{Jing-Wu~Li$^{1}$\footnote{Electronic address: {\Blue{lijw@jsnu.edu.cn}}}}

\author{Zhi-Tian~Zou$^{3}$\footnote{Electronic address: {\Blue{zouzt@ytu.edu.cn}}}}


\affiliation{
$^1$ School of Physics and Electronic Engineering, Jiangsu Normal University, Xuzhou, Jiangsu 221116,
People's Republic of China\\
$^2$ Department of Physics and Institute of Theoretical
Physics, Nanjing Normal University, Nanjing, Jiangsu 210023,
People's Republic of China\\
$^3$ Department of Physics, Yantai University, Yantai, Shandong 264005, People's Republic of China}


\date{\today}

\begin{abstract}

We study 20 
charmless hadronic $B \to f_1 P$ decays
in the perturbative QCD(pQCD) formalism
with $B$ denoting $B_u$, $B_d$, and $B_s$ mesons;
$P$ standing for the light pseudoscalar mesons;
and $f_1$ representing axial-vector mesons $f_1(1285)$
and $f_1(1420)$
that result from a mixing of quark-flavor
$f_{1q}[\frac{u\bar u+ d\bar d}{\sqrt{2}}]$ and
$f_{1s}[s\bar s]$ states with the angle $\phi_{f_1}$.
The estimations of \cp-averaged branching ratios and \cp asymmetries
of the considered $B \to f_1 P$ decays, in which the $B_s \to f_1 P$ modes
are investigated for the first time,
are presented in the pQCD approach with $\phi_{f_1} \sim 24^\circ$
from recently
measured $B_{d/s} \to J/\psi f_1(1285)$ decays.
It is found that
(a) the tree(penguin) dominant $B^+ \to f_1 \pi^+(K^+)$
decays with large branching ratios[${\cal O}(10^{-6})$]
and large direct \cp violations(around $14\% \sim 28\%$ in magnitude)
simultaneously are believed to be clearly measurable at the LHCb and Belle II 
experiments;
(b) the $B_d \to f_1 K_S^0$ and $B_s \to f_1 (\eta, \eta^{\prime})$ decays
with nearly pure penguin contributions and safely negligible tree pollution also
have large decay rates in the order of $10^{-6} \sim 10^{-5}$, which can be
confronted with the experimental measurements in the near future;
(c) as the alternative channels, the $B^+ \to f_1 (\pi^+, K^+)$ and $B_d \to f_1 K_S^0$
decays have the supplementary power in providing more effective constraints on the
Cabibbo-Kobayashi-Maskawa weak phases $\alpha$, $\gamma$, and $\beta$, correspondingly,
which are explicitly analyzed through the large decay rates and the direct and mixing-induced \cp
asymmetries in the pQCD approach and are expected to be stringently examined
by the measurements with high precision;
(d) the weak annihilation amplitudes play important roles in the $B^+ \to f_1(1420) K^+$,
$B_d \to f_1(1420) K_S^0$, $B_s \to f_1(1420) \eta^{\prime}$ decays, and so on, which would
offer more evidence, once they are confirmed by the experiments,
to identify the soft-collinear effective theory and the pQCD approach
on the evaluations of annihilation diagrams and to
help further understand the annihilation mechanism in the heavy $B$ meson decays;
(e) combined with the future precise tests, the considered decays can provide more information
to further understand the mixing angle $\phi_{f_1}$ and the nature of the $f_1$ mesons
in depth after the confirmations on the reliability of the pQCD calculations in the present work.

\end{abstract}

\maketitle

%
%

\section{Introduction}

It is well known that nonleptonic weak decays of heavy
$B$(specifically, $B_u$, $B_d$, $B_s$, and $B_c$) mesons
can not only provide the important information to search for \cp violation and
further constrain the Cabibbo-Kobayashi-Maskawa(CKM) parameters in the
standard model(SM), but also reveal the deviations from the SM, i.e.,
the signals of exotic new physics beyond the SM. Furthermore, comparison
of theoretical predictions and experimental data for the physical observables
may also help us understand the hadronic structure of the involved bound states
deeply~\cite{Feldmann:2014iha}.  In contrast to the
traditional $B \to PP, PV$, and $VV$ decays, the alternative channels such
as $B \to AP$ decays ($A$ is the axial-vector meson) to be largely detected
in experiments in the near future
may give additional and complementary information on exclusive nonleptonic
weak decays of heavy $B$ mesons~\cite{Calderon:2007nw}; e.g., due to $V_{tb}^*V_{ts} =
- V_{cb}^* V_{cs} [1+ {\cal O}(\lambda^2)]$,
the $b \to s q\bar q$ penguin-dominated decays have the same CKM phase as the
$b \to c \bar c s$ tree level decays~\cite{Agashe:2014kda}.
Therefore, the $b \to s q \bar q$ mediated $B \to AP$ decays such as $B^0 \to a_1(1260)[b_1(1235)] K_S^0$
$\pi K_1(1270)[K_1(1400)]$, $f_1 K_S^0$ etc. can provide $\sin2\beta$ measurements ($\beta$ is the CKM weak phase) in the SM complementarily.

Very recently, the Large Hadron Collider beauty(LHCb) Collaboration
reported the first measurements of $B_{d/s} \to J/\psi f_1(1285)$
decays~\cite{Aaij:2013rja}, where the final state $f_1(1285)$ was observed for the
first time in heavy $B$ meson decays.
In the conventional two quark structure, $f_1(1285)$ and its partner
$f_1(1420)$~\cite{Close:1997nm,Li:2000dy} [hereafter, for the sake of simplicity,
we will use $f_1$ to denote both $f_1(1285)$ and $f_1(1420)$ unless otherwise stated]
are considered as the orbital
excitation of the $q\bar q$ system, specifically, the light $p$-wave axial-vector
flavorless mesons. In terms of the spectroscopic notation $^{(2S+1)}L_J$
with ${\bf J} ={\bf L} + {\bf S}$,
both $f_1$ mesons belong to $^3\!P_1$ nonet carrying the quantum
number $J^{PC} = 1^{++}$~\cite{Agashe:2014kda}.
Similar to the $\eta-\eta^{\prime}$ mixing in the pseudoscalar
sector~\cite{Agashe:2014kda}, these two $f_1$ mesons are believed to
be a mixture resulting from the mixing
between nonstrange $f_{1q} \equiv (u\bar u+ d\bar d)/\sqrt{2}$ and
strange $f_{1s}\equiv s\bar s$ states in the popular quark-flavor basis
with a single mixing angle $\phi_{f_1}$. And for the mixing angle $\phi_{f_1}$,
there are several explorations that have been performed from theory and experiment
sides. However, the value of $\phi_{f_1}$ is still in controversy
presently (see Ref.~\cite{Liu:2014doa} and references therein).
It is necessary to point out that the mixing angle $\phi_{f_1}$ has important roles
in investigating the properties of $f_1$ mesons themselves, but also of
strange axial-vector $K_1$ mesons, i.e., $K_1{1270}$ and $K_1(1400)$, by
constraining the mixing angle $\theta_{K_1}$ between two distinct types of
axial-vector $K_{1A}(^3\!P_1)$ and $K_{1B}(^1\!P_1)$ states. The underlying reason
is that when the $f_1(1285)-f_1(1420)$ mixing angle $\phi_{f_1}$ is determined
from the mass relations related with the masses of $K_{1A}$ and $K_{1B}$,
it eventually depends on the mixing angle
$\theta_{K_1}$~\cite{Cheng:2011pb}.
With the successful running of LHC and the forthcoming Belle II experiments, 
it is therefore expected that these first observations of
$B_{d/s} \to J/\psi f_1(1285)$ decays will motivate the people to
explore the mixing angle $\phi_{f_1}$ and the properties of both
$f_1$ mesons in more relevant $B$ meson decay processes at both
experimental and theoretical aspects. Of course, in view of some of the
axial-vector mesons such as $a_1(1260)$ and $K_1$ that have been seen in
two-body hadronic $D$ meson decays~\cite{Agashe:2014kda}, it is also believed
that the information on both $f_1$ mesons could be obtained from heavy
$c$-quark decays in the near future.

In this work, we will therefore study 20 charmless hadronic
$B \to f_1 P$\footnote{In the literature
\cite{Liu:2010da}, two of us(X.~Liu and Z.J.~Xiao) have studied the $B_c \to f_1 P$
decays occurring only via the pure annihilation diagrams in the SM
within the framework of the perturbative QCD(pQCD) factorization approach~\cite{Keum:2000ph,Lu:2000em,Li:2003yj}.}
decays, in which $B$ stands for
$B_u$, $B_d$, and $B_s$, respectively, and $P$ denotes the
light pseudoscalar pion, kaon, and $\eta$ and $\eta'$ mesons.
From the experimental
point of view, up to now, only two penguin-dominated $B^+ \to f_1 K^+$ decays have been measured by the
{\it BABAR} Collaboration in 2007~\cite{Burke:2007zz}. The preliminary upper limits
on the decay rates have been placed at the 90\% confidence level as~\cite{Agashe:2014kda}
\beq
Br(B^+ \to f_1(1285) K^+) &<&   2.0 \times 10^{-6}\;,
\label{eq:exp-br-1285}
\eeq
for $B^+ \to f_1(1285) K^+$ decay and
\beq
Br(B^+ \to f_1(1420) K^+)\cdot Br(f_1(1420) \to \bar K^* K) &<&
4.1 \times 10^{-6}\;,
\label{eq:exp-br-1420-1}   \\
Br(B^+ \to f_1(1420) K^+)\cdot Br(f_1(1420) \to \eta \pi\pi) &<&
2.9 \times 10^{-6}\;,
\label{eq:exp-br-1420-2}
\eeq
for $B^+ \to f_1(1420) K^+$ decay, respectively. However,
due to the lack of the information on the decay rates of
$f_1(1420) \to \bar K^* K$ and $f_1(1420) \to \eta \pi\pi$ decays,
 the upper limits of $Br(B^+ \to f_1(1420) K^+)$ cannot be
extracted effectively. But, this status will be greatly improved
in present and future experiments, notably at running LHCb
and forthcoming Belle II. 
Also, other $B \to f_1 P$ decays
are expected to be detected with good precision
at the relevant experiments in the near future.

From the theoretical point of view, to our best knowledge,
G.~Calder${\rm \acute{o}}$n {\it et al.}
have carried out the calculations of $B_{u,d} \to f_1 P$ decays
in the framework of naive factorization
with the form factors of $B \to f_1$ obtained in the improved Isgur-
Scora-Grinstein-Wise quark model~\cite{Calderon:2007nw},
while Cheng and Yang have studied the
decay rates and direct \cp asymmetries of $B_{u,d} \to f_1 (\pi, K)$
modes within the framework of QCD factorization (QCDF)
with the form factors evaluated in the QCD sum rule~\cite{Cheng:2007mx}.
Note that the $B_s \to f_1 P$ decays have never
been studied yet in any methods or approaches up to this date.
And, it should be stressed that the predictions
of the branching ratios for $B_{u,d} \to f_1 P$ decays in naive factorization
are so crude that we cannot make effective comparison
for relevant $B \to f_1 P$ modes.
For $B^+ \to f_1 K^+$ decays for example, on one hand, the authors did not specify
$f_1(1285)$ and $f_1(1420)$~\cite{Calderon:2007nw},
which then could not provide effectively the useful
information on the mixing angle $\phi_{f_1}$ from these considered decays; on
the other hand, as discussed in Ref.~\cite{Cheng:2007mx}, the $^3\!P_1$ meson
behaves analogously to the vector meson, it is then naively expected that
$Br(B^+ \to f_1(1285) K^+) \sim Br(B^+ \to \omega K^+)$ and $Br(B^+ \to f_1(1420) K^+) \sim
Br(B^+ \to \phi K^+)$ if $f_1(1285)$ and $f_1(1420)$ are significantly dominated by the
$f_{1q}$ and $f_{1s}$ components, respectively. Furthermore, in principle, in view of the
$f_1(1285)-f_1(1420)$ mixing, the branching ratios of $B^+ \to (f_1(1285), f_1(1420)) K^+$ are generally
a bit smaller than those of $B^+ \to (\omega, \phi) K^+$ ones correspondingly.
However, the branching ratio of $B^+ \to f_1 K^+$ predicted
in the naive factorization is around $3 \times 10^{-5}$, which is much larger
than that of the corresponding $VP$ modes, i.e., $B^+ \to \omega K^+$ and
$B^+ \to \phi K^+$~\cite{Agashe:2014kda}.  As for the investigations
of $B^+ \to f_1 K^+ $ decays in the QCDF approach~\cite{Cheng:2007mx},
the authors specified $f_1(1285)$ and $f_1(1420)$ and
considered their mixing with two different sets of angles, $\theta_{^3\!P_1} \sim 27.9^\circ$
and $53.2^\circ$, 
in the flavor singlet-octet basis.
And the decay rates are barely consistent with the preliminary upper limits within
very large errors. However, the pattern exhibited from the numerical results with
$\theta_{^3\!P_1} \sim 53.2^\circ$ is more favored by the available upper limits.
As aforementioned, because of the similar behavior between the vector meson
and $^3\!P_1$ axial-vector meson, the relation
$Br(B^+ \to f_1(1285) K^+) < Br(B^+ \to f_1(1420) K^+)$
is expected to be highly preferred, as it should be.

In order to collect more information on the nature of both $f_1$ mesons and
further understand the heavy flavor $B$ physics,
we will study the physical observables such as \cp-averaged branching ratios
and \cp-violating asymmetries of 20 charmless hadronic $B \to f_1 P$ decays
by employing the low energy
effective Hamiltonian~\cite{Buchalla:1995vs}
and the pQCD approach~\cite{Keum:2000ph,Lu:2000em,Li:2003yj} based on the $k_T$ factorization theorem.
Though some efforts have been made on the next-to-leading order pQCD formalism~\cite{Li:2010nn,Cheng:2014gba},
we here will still consider the perturbative evaluations at leading order,
which are believed to be the dominant contributions perturbatively.
As is well known, the pQCD approach
is free of end-point singularity
and the Sudakov formalism makes it more self-consistent by keeping
the transverse momentum $k_T$ of the quarks.
More importantly, as the well-known advantage of the pQCD approach,
 we can explicitly calculate the weak annihilation types
of diagrams without any parametrizations,
apart from the traditional
factorizable and nonfactorizable emission ones, though
a different viewpoint on the evaluations and the magnitudes\footnote{As a matter of fact, the recent works~\cite{Zhu:2011mm,Chang:2014rla} in the
framework of QCDF confirmed that there should exist complex
annihilation contributions with large imaginary parts
in $B_{u,d,s} \to \pi\pi, \pi K, KK$
decays  by fits to the experimental data, which support the concept on the calculations
 of the annihilation diagrams in the pQCD approach~\cite{Chay:2007ep} to some extent.}
of weak annihilation contributions has been proposed in the soft-collinear
effective theory~\cite{Arnesen:2006vb}.
It is worth stressing that the pQCD predictions on the
annihilation contributions in the heavy $B$ meson decays have been
tested at various aspects,
 e.g., see Refs.
~\cite{Lu:2002iv,Li:2004ep,Ali:2007ff,Xiao:2011tx,Keum:2000ph,Lu:2000em,Hong:2005wj}.
Typically, for example, the evaluations
of the pure annihilation $B_d \to K^+ K^-$ and $B_s \to \pi^+ \pi^-$
decays in the pQCD approach~\cite{Li:2004ep,Ali:2007ff,Xiao:2011tx}
are in good consistency with the recent
measurements by both CDF and LHCb Collaborations~\cite{Aaltonen:2011jv,Ruffini:2013jea,Aaij:2012as}.
Therefore, the weak annihilation contributions to the considered $B \to f_1 P$ decays
will be explicitly analyzed in this work, which are expected to be helpful to understand the
annihilation mechanism in the heavy $B$ meson decays.

The paper is organized as follows. In Sec.~\ref{sec:form}, we present
the formalism, hadron wave functions and perturbative calculations
of the considered 20 $B \to f_1 P$ decays in the pQCD approach.
The numerical results and the corresponding phenomenological
analyses are addressed in Sec.~\ref{sec:randd}. Finally,
Sec.~\ref{sec:summary} contains the main conclusions and a short
summary.

%
%
\section{ Formalism and Perturbative Calculations}\label{sec:form}

For the considered $B \to f_1 P$ decays,
the related weak effective
Hamiltonian $H_{{\rm eff}}$~\cite{Buchalla:1995vs} can be written as
\beq
H_{\rm eff}\, &=&\, {G_F\over\sqrt{2}}
\biggl\{ V^*_{ub}V_{uD} [ C_1(\mu)O_1^{u}(\mu)
+C_2(\mu)O_2^{u}(\mu) ]
 - V^*_{tb}V_{tD} [ \sum_{i=3}^{10}C_i(\mu)O_i(\mu) ] \biggr\}+ {\rm H.c.}\;,
\label{eq:heff}
\eeq
with $D$ the light down-type quark $d$ or $s$, the Fermi constant $G_F=1.16639\times 10^{-5}{\rm
GeV}^{-2}$, 
CKM matrix elements $V$,
and Wilson coefficients $C_i(\mu)$ at the renormalization scale
$\mu$. The local four-quark
operators $O_i(i=1,\cdots,10)$ are written as
\begin{enumerate}
\item[]{(1) current-current(tree) operators}
\begin{eqnarray}
{\renewcommand\arraystretch{1.5}
\begin{array}{ll}
\displaystyle
O_1^{u}\, =\,
(\bar{D}_\alpha u_\beta)_{V-A}(\bar{u}_\beta b_\alpha)_{V-A}\;,
& \displaystyle
O_2^{u}\, =\, (\bar{D}_\alpha u_\alpha)_{V-A}(\bar{u}_\beta b_\beta)_{V-A}\;;
\end{array}}
\label{eq:operators-1}
\end{eqnarray}

\item[]{(2) QCD penguin operators}
\begin{eqnarray}
{\renewcommand\arraystretch{1.5}
\begin{array}{ll}
\displaystyle
O_3\, =\, (\bar{D}_\alpha b_\alpha)_{V-A}\sum_{q'}(\bar{q}'_\beta q'_\beta)_{V-A}\;,
& \displaystyle
O_4\, =\, (\bar{D}_\alpha b_\beta)_{V-A}\sum_{q'}(\bar{q}'_\beta q'_\alpha)_{V-A}\;,
\\
\displaystyle
O_5\, =\, (\bar{D}_\alpha b_\alpha)_{V-A}\sum_{q'}(\bar{q}'_\beta q'_\beta)_{V+A}\;,
& \displaystyle
O_6\, =\, (\bar{D}_\alpha b_\beta)_{V-A}\sum_{q'}(\bar{q}'_\beta q'_\alpha)_{V+A}\;;
\end{array}}
\label{eq:operators-2}
\end{eqnarray}

\item[]{(3) electroweak penguin operators}
\begin{eqnarray}
{\renewcommand\arraystretch{1.5}
\begin{array}{ll}
\displaystyle
O_7\, =\,
\frac{3}{2}(\bar{D}_\alpha b_\alpha)_{V-A}\sum_{q'}e_{q'}(\bar{q}'_\beta q'_\beta)_{V+A}\;,
& \displaystyle
O_8\, =\,
\frac{3}{2}(\bar{D}_\alpha b_\beta)_{V-A}\sum_{q'}e_{q'}(\bar{q}'_\beta q'_\alpha)_{V+A}\;,
\\
\displaystyle
O_9\, =\,
\frac{3}{2}(\bar{D}_\alpha b_\alpha)_{V-A}\sum_{q'}e_{q'}(\bar{q}'_\beta q'_\beta)_{V-A}\;,
& \displaystyle
O_{10}\, =\,
\frac{3}{2}(\bar{D}_\alpha b_\beta)_{V-A}\sum_{q'}e_{q'}(\bar{q}'_\beta q'_\alpha)_{V-A}\;.
\end{array}}
\label{eq:operators-3}
\end{eqnarray}
\end{enumerate}
with the color indices $\alpha, \ \beta$(not to be confused with the CKM weak phases
$\alpha$ and $\beta$) and the notations
$(\bar{q}'q')_{V\pm A} = \bar q' \gamma_\mu (1\pm \gamma_5)q'$.
The index $q'$ in the summation of the above operators runs
through $u,\;d,\;s$, $c$, and $b$. The standard combinations
$a_i$ of the Wilson coefficients $C_i$ are defined as follows:
\beq
a_1 &=& C_2 + \frac{C_1}{3}\;, \qquad a_2 = C_1 + \frac{C_2}{3} \;,
\qquad a_i= C_i+C_{i\pm 1}/3,\quad  i=3-10.
\eeq
where $C_2 \sim 1$ is
the largest one among all Wilson coefficients and the upper (lower)
sign applies, when $i$ is odd (even). It is noted that, though
the next-to-leading order Wilson coefficients have already been available~\cite{Buchalla:1995vs},
we will still adopt the leading order ones to keep the consistency, since the short distance
contributions of the considered decays are calculated at leading order[${\cal O}(\alpha_s)$]
in the pQCD approach. This is also a consistent way to
cancel the explicit $\mu$ dependence in the theoretical formulas.
For the renormalization group evolution of the Wilson coefficients
from higher scale to lower scale, the expressions are directly
taken from Ref.~\cite{Lu:2000em}.

Nowadays, the pQCD approach has been one of the popular factorization
methods based on the QCD theory
to evaluate the hadronic matrix elements in the heavy $B$ meson decays.
The unique point of the pQCD approach is that it keeps
the transverse momentum $k_T$,
which will act as an infrared regulator and smear the end-point
singularity when the quark momentum fraction $x$ approaches 0, of the valence quarks
in the calculation of the hadronic matrix elements.
Then, all the $B$ meson transition form factors, non-factorizable spectator and
annihilation contributions are calculable in the framework
of the $k_T$ factorization.
The decay amplitude of $B \to f_1 P$ decays in the pQCD approach
can be conceptually written as
\beq
{\cal A}(B \to f_1 P) &\sim &\int\!\! d^4k_1d^4k_2d^4k_3 \mathrm{Tr} \left [
C(t) \Phi_{B}(k_1) \Phi_{f_1}(k_2) \Phi_{P}(k_3)
H(k_1,k_2,k_3, t) \right ],\label{eq:a1}
\eeq
where
$k_i$'s are the momenta of (light) quarks in the initial and final states,
$\mathrm{Tr}$ represents the trace over Dirac and color indices, and
$C(t)$ is the Wilson coefficient which results from the radiative corrections at
short distance. In the above convolution, $C(t)$ includes the
harder dynamics at larger scale than $m_{B}$ scale and describes
the evolution of local $4$-Fermi operators from $m_W$ (the $W$
boson mass) down to $t\sim\mathcal{O}(\sqrt{\Lambda_{\rm QCD}
m_{B}})$ scale, where $\Lambda_{\rm QCD}$ is the hadronic scale.
The $\Phi$ stands for the wave function describing
hadronization of the quark and antiquark to the meson, which
is independent of the specific processes and usually determined by
employing nonperturbative QCD techniques such as lattice QCD(LQCD)
or other well-measured processes.
The function $H(k_1,k_2,k_3,t)$ describes the four-quark operator
and the spectator quark connected by
 a hard gluon with the hard intermediate scale
 $\mathcal{O}(\sqrt{\Lambda_{\rm QCD} m_{B}})$.
Therefore, this hard part $H$ can be calculated perturbatively.

Since the $b$ quark is rather heavy, we thus work in the frame with the
$B$ meson at rest for simplicity, i.e., with the $B$ meson momentum
$P_1=\frac{m_{B}}{\sqrt{2}}(1,1,{\bf 0}_T)$ in the light-cone
coordinate. For the considered $B \to f_1 P$ decays, it is assumed that
the $f_1$ and $P$ mesons move in the plus and minus $z$
direction carrying the momentum $P_2$ and $P_3$, respectively.
Then the momenta of the two final states can be
written as
\beq
     P_2 =\frac{m_{B}}{\sqrt{2}} (1,r_{f_1}^2,{\bf 0}_T), \quad
     P_3 =\frac{m_{B}}{\sqrt{2}} (0,1-r_{f_1}^2,{\bf 0}_T),
\eeq
respectively, where $r_{f_1}=m_{f_1}/m_{B}$ and the masses of light
pseudoscalar pion, kaon, and $\eta$ and $\eta'$ have been neglected.
For the axial-vector meson $f_1$, its longitudinal polarization vector $\epsilon_2^L=\frac{m_{B}}{\sqrt{2}m_{f_1}} (1,
-r_{f_1}^2,{\bf 0}_T)$.
By choosing the quark momenta in $B$, $f_1$ and $P$ mesons as $k_1$,
$k_2$, and $k_3$, respectively, and defining
\beq
k_1 = (x_1P_1^+,0,{\bf k}_{1T}), \quad k_2 = x_2 P_2+(0,0,{\bf k}_{2T}),
\quad k_3 = x_3 P_3+(0, 0,{\bf k}_{3T}).
\eeq
then, the
integration over $k_1^-$, $k_2^-$, and $k_3^+$ in Eq.~(\ref{eq:a1})
will give the more explicit form of decay amplitude in the pQCD approach,
\beq
{\cal A}(B \to f_1\;P) &\sim &\int\!\! d x_1 d x_2 d x_3 b_1
d b_1 b_2 d b_2 b_3 d b_3 \non && \cdot \mathrm{Tr} \left [ C(t)
\Phi_{B}(x_1,b_1) \Phi_{f_1}(x_2,b_2) \Phi_{P}(x_3, b_3) H(x_i,
b_i, t) S_t(x_i)\, e^{-S(t)} \right ]\;
\label{eq:a2}
\eeq
where $b_i$ is the conjugate space coordinate of $k_{iT}$, and $t$ is the
largest running scale in the hard kernel $H(x_i,b_i,t)$. The large
logarithms $\ln (m_W/t)$ are included in the Wilson coefficients
$C(t)$. Note that $S_t(x_i)$ and $e^{-S(t)}$ are the two important elements in the
perturbative calculations with the pQCD approach.
The former is a jet function from threshold resummation,
which can strongly suppress the behavior in the small $x$
region~\cite{Li:2001ay,Li:2002mi};
while the latter is a Sudakov factor from $k_T$ resummation,
which can effectively suppress the soft dynamics in
the small $k_T$ region~\cite{Botts:1989kf,Li:1992nu}.
These resummation effects therefore guarantee the removal of
the end-point singularities. Thus it makes the
perturbative calculation of the hard part $H$ applicable at
intermediate scale.
We will calculate
analytically the function $H(x_i,b_i,t)$ for the $B \to f_1 P$ decays
at LO in the $\alpha_s$ expansion and give the convoluted
amplitudes in the next section.

The heavy $B$ meson is usually treated as a heavy-light system and its light-cone wave function
can generally be defined as~\cite{Keum:2000ph,Lu:2000em,Lu:2002ny}
\beq
\Phi_{B}
&=& \frac{i}{\sqrt{2N_c}}
\left\{(\psl +m_{B})\gamma_5
 \phi_{B}(x, k_T) \right\}_{\alpha\beta}\;;
\label{eq:def-bq}
\eeq
in which
$\alpha,\beta$ are the color indices; 
$P(m)$ is the momentum(mass) of the $B$ meson; $N_c$ is the color factor; and
$k_T$ is the intrinsic transverse momentum
of the light quark in the $B$ meson.

In Eq.~(\ref{eq:def-bq}),
$\phi_{B}(x,k_T)$ is the $B$ meson distribution amplitude,
which satisfies the following normalization condition,
\beq
\int_0^1 dx \phi_{B}(x, b=0) &=& \frac{f_{B}}{2 \sqrt{2N_c}}\;,\label{eq:norm}
\eeq
where $b$ is the conjugate space coordinate of transverse momentum $k_T$ and $f_B$
is the decay constant of the $B$ meson.

For the pseudoscalar $P$ meson, the light-cone wave function can generally be
defined as~\cite{Chernyak:1983ej,Ball:1998tj}
\beq
\Phi_P(x) &=& \frac{i}{\sqrt{2 N_c}} \gamma_5
\left\{\psl \phi_P^A(x)+ m_0^P \phi_P^P(x) + m_0^P (\nsl \vsl -1)
\phi_P^T(x)\right\}_{\alpha\beta}
\eeq
where $\phi_P^{A}$ and $\phi_P^{P,T}$ are the twist-2 and twist-3 distribution amplitudes,
and $m_0^P$ is the chiral enhancement factor of the meson,
while $x$ denotes the momentum
fraction carried by quark in the meson and $n=(1,0,{\bf 0}_T)$
and $v=(0,1,{\bf 0}_T)$ are the dimensionless lightlike unit vectors.

The light-cone wave function of the axial-vector $f_1$ mesons
has been given in the QCD sum rule as~\cite{Yang:2007zt,Li:2009tx}
 \beq
\Phi^{L}_{f_{1}}
 &=&  \frac{1}{\sqrt{2 N_c}}
 \gamma_5  \biggl\{ m_{f_{1}}\, {\epsl}_L \,\phi_{f_{1}}(x)  +
 {\epsl}_L \, \psl\,\phi_{f_{1}}^t(x)  + m_{f_{1}}\, \phi_{f_{1}}^s(x) \biggr\}_{\alpha\beta}\;,
 \eeq
for longitudinal polarization
with the polarization vector $\epsilon_L$,
satisfying $P \cdot \epsilon=0$, where $\phi_{f_1}$(not to be confused with the angle
$\phi_{f_1}$ in the mixing of $f_1$ mesons) and $\phi_{f_1}^{s,t}$ are the twist-2
and twist-3 distribution amplitudes, respectively.
All the explicit forms of the aforementioned hadronic distribution amplitudes
in the considered $B \to f_1 P$ decays
can be seen in the Appendix.

\begin{figure}[!!htb]
\centering
\begin{tabular}{l}
\includegraphics[width=0.8\textwidth]{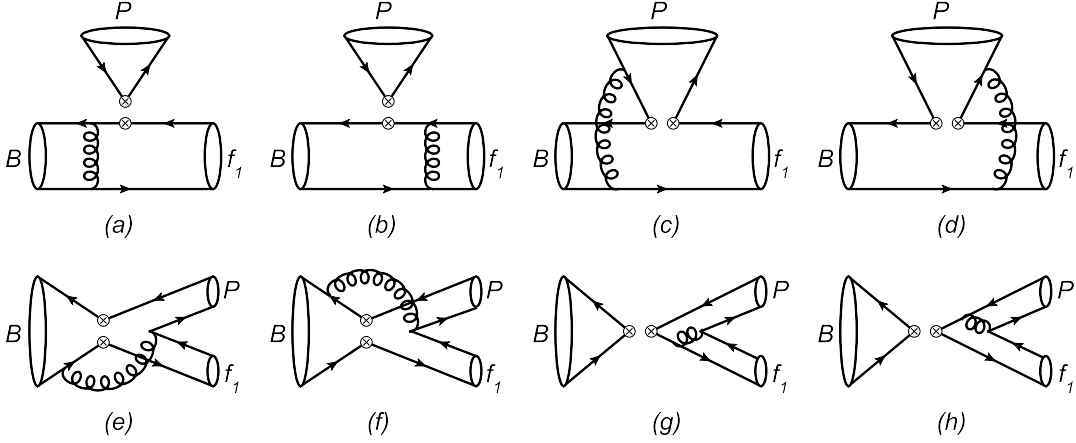}
\end{tabular}
\caption{Typical Feynman diagrams contributing to
 $B \to P f_1$ decays in the pQCD approach
at leading order, in which $P$ denotes the light
pseudoscalar $\pi$, $K$, and $\eta$ and $\eta'$ mesons
and
$f_1$ stands for the axial-vector
$f_1(1285)$ and $f_1(1420)$, respectively. When we
exchange the positions of $P$ and $f_1$, the other
eight diagrams contributing to the considered decays
will be easily obtained.}
  \label{fig:fig1}
\end{figure}

Now we come to the analytically perturbative calculations of the factorization formulas
for the $B \to f_1 P$ decays in the pQCD approach.
From the effective Hamiltonian (\ref{eq:heff}),
there are eight types of diagrams contributing to the $B
\to P f_1$ decays as illustrated in Fig.~\ref{fig:fig1}.
For the factorizable emission($fe$) diagrams, with Eq.~(\ref{eq:a2}),
the analytic expressions of the decay amplitudes from
different operators are given as follows:
\begin{itemize}
\item $(V-A)(V-A)$ operators:
 \beq
F_{fe}&=& -8 \pi C_F f_{P} m_{B}^2 \int_0^1 d x_{1} dx_{3}\,
\int_{0}^{\infty} b_1 db_1 b_3 db_3\,
 \phi_{B}(x_1,b_1)\left\{ \left[(1+x_3 )\phi_{f_1}(x_3)
 + r_{f_1} (1-2 x_3)
\right. \right. \non && \left. \left. \times
(\phi^s_{f_1}(x_3)+\phi^t_{f_1}(x_3))\right]
h_{fe}(x_1,x_3,b_1,b_3)E_{fe}(t_a)+ 2\; r_{f_1}\; \phi^s_{f_1} (x_3)\;
h_{fe}(x_3,x_1,b_3,b_1) \;E_{fe}(t_b)
\right\}\;, \label{eq:pf1-fe}
 \eeq
\item $(V-A)(V+A)$ operators:
\beq
F_{fe}^{P1}&=& - F_{fe}\;, \label{eq:pf1-fe-p1}
\eeq
\item $(S-P)(S+P)$ operators:
 \beq
 F_{fe}^{P2}&=& -16\pi C_F f_{P}
m_{B}^2 r_0^P  \int_0^1 d x_{1} dx_{3}\, \int_{0}^{\infty} b_1 db_1 b_3
db_3\,   \phi_{B}(x_1,b_1)\left\{
\left[\phi_{f_1}(x_3)+ r_{f_1} \left[(2+x_3) \phi^s_{f_1}(x_3) \right.\right.\right.
\non && \left. \left. \left.- x_3
\phi^t_{f_1}(x_3)\right]\right]h_{fe}(x_1,x_3,b_1,b_3)E_{fe}(t_a)+ 2\; r_{f_1}\;
\phi^s_{f_1} (x_3) \; h_{fe}(x_3,x_1,b_3,b_1)\;
E_{fe}(t_b)\right\} \label{eq:pf1-fe-p2}\;;
\eeq
\end{itemize}
where $r_0^P=m_{0}^P/m_{B}$ and $C_F=4/3$ is a color
factor. The convolution functions $E_i$, the running hard
scales $t_i$, and the hard functions $h_i$ can be referred to in
Ref.~\cite{Liu:2005mm}.

For the nonfactorizable emission($nfe$) diagrams in Figs.~\ref{fig:fig1}(c)
and \ref{fig:fig1}(d), the corresponding
decay amplitudes can be written as
\begin{itemize}
\item $(V-A)(V-A)$ operators:
\beq
M_{nfe}&=&-
\frac{32}{\sqrt{6}}\pi C_F m_{B}^2 \int_{0}^{1}d x_{1}d x_{2}\,d
x_{3}\,\int_{0}^{\infty} b_1d b_1 b_2d b_2\,
\phi_{B}(x_1,b_1) \phi_{P}^A(x_2)
\non && \times
\left\{ \left[(1-x_2)\phi_{f_1}(x_3)
 - r_{f_1} x_3 \left(\phi^s_{f_1}(x_3)-
\phi^t_{f_1}(x_3)\right)\right]
E_{nfe}(t_c)h_{nfe}^c(x_1,x_2,x_3,b_1,b_2) \right. \non && \left.
-\left[(x_2+x_3)\phi_{f_1}(x_3) - r_{f_1} x_3 (\phi^s_{f_1}(x_3)+
\phi^t_{f_1}(x_3))\right] E_{nfe}(t_d)h_{nfe}^d(x_1,x_2,x_3,b_1,b_2)
\right\} \label{eq:pf1-nfe}\; ,
\eeq
\item $(V-A)(V+A)$ operators:
\beq
M_{nfe}^{P1}&=&-\frac{32}{\sqrt{6}}\pi C_F m_{B}^2 \int_{0}^{1}d
x_{1}d x_{2}\,d x_{3}\,\int_{0}^{\infty} b_1d b_1 b_2d b_2\,
\phi_{B}(x_1,b_1) r_{0}^P  \left\{
\left[(1-x_2)(\phi_{P}^P(x_2)
+\phi_{P}^T(x_2)) \phi_{f_1}(x_3)- r_{f_1} \right.\right.\non
 & & \left.\left. \times \left((1-x_2-x_3)\left(\phi_P^P(x_2)\phi_{f_1}^t(x_3)
 -\phi_P^T(x_2)\phi_{f_1}^s(x_3)\right)-(1-x_2+x_3)\left(\phi_P^P(x_2)\phi_{f_1}^s(x_3)
 -\phi_P^T(x_2)\phi_{f_1}^t(x_3)\right) \right)\right]
\right. \non &&\left. \times
E_{nfe}(t_c) h_{nfe}^c(x_1,x_2,x_3,b_1,b_2) -h_{nfe}^d(x_1,x_2,x_3,b_1,b_2)E_{nfe}(t_d)
\left[x_2\left.(\phi_{P}^P(x_2)-\phi_{P}^T(x_2)\right)\phi_{f_1}(x_3)
\right.\right.\non &&\left.\left. + r_{f_1}
(x_2 (\phi_{P}^P(x_2)-\phi_{P}^T(x_2))(\phi^s_{f_1}(x_3)- \phi^t_{f_1}(x_3)) + x_3
(\phi_{P}^P(x_2)+\phi_{P}^T(x_2))(\phi^s_{f_1}(x_3)+
\phi^t_{f_1}(x_3)))\right]
\right\} \label{eq:pf1-nfe-p1}\;,
\eeq
\item $(S-P)(S+P)$ operators:
\beq
 M_{nfe}^{P2}&=& -\frac{32}{\sqrt{6}}\pi C_F m_{B}^2
\int_{0}^{1}d x_{1}d x_{2}\,d x_{3}\,\int_{0}^{\infty} b_1d b_1 b_2d
b_2\, \phi_{B}(x_1,b_1)  \phi_{P}^A(x_2)  \non && \times
\left\{\left[(x_2-x_3-1)\phi_{f_1}(x_3) +
r_{f_1} x_3 (\phi^s_{f_1}(x_3)+
\phi^t_{f_1}(x_3))\right] E_{nfe}(t_c)h_{nfe}^c(x_1,x_2,x_3,b_1,b_2)
\right. \non
& &\left.
+\left[ x_2 \phi_{f_1}(x_3) - r_{f_1} x_3
(\phi^s_{f_1}(x_3)- \phi^t_{f_1}(x_3))\right]
h_{nfe}^d(x_1,x_2,x_3,b_1,b_2) E_{nfe}(t_d) \right\} \label{eq:pf1-nfe-p2}\;;
\eeq
\end{itemize}

The Feynman diagrams shown in Figs.~\ref{fig:fig1}(e) and \ref{fig:fig1}(f) are
the nonfactorizable annihilation($nfa$) ones, whose contributions are
\begin{itemize}
\item $(V-A)(V-A)$ operators:
\beq
M_{nfa}&=& -\frac{32}{\sqrt{6}}\pi C_F m_{B}^2
\int_{0}^{1}d x_{1}d x_{2}\,d x_{3}\,\int_{0}^{\infty} b_1d b_1 b_2d
b_2\,  \phi_{B}(x_1,b_1)\left\{
\left[(1-x_3)\phi_{P}^A(x_2)\phi_{f_1}(x_3)+ r_0^P  r_{f_1}
\left(\phi_{P}^P(x_2) \right.\right.\right.
\non && \left.\left.\left.
\times [(1+x_2-x_3) \phi^s_{f_1}(x_3) -
(1-x_2-x_3)\phi^t_{f_1}(x_3)]+\phi_{P}^T(x_2)
 \left[(1-x_2-x_3)\phi_{f_1}^s(x_3)-(1+x_2-x_3)\right.\right.\right.\right.\non
& & \left.\left.\left.\left. \times \phi_{f_1}^t(x_3)\right]
\right)\right] E_{nfa}(t_e)   h_{nfa}^e(x_1,x_2,x_3,b_1,b_2)-E_{nfa}(t_f)
h_{nfa}^f(x_1,x_2,x_3,b_1,b_2) \left[x_2 \phi_{P}^A(x_2)\phi_{f_1}(x_3)
\right.\right. \non && \left.\left. + r_0^P r_{f_1}
\left(\phi_{P}^P(x_2)[(x_2-x_3+3) \phi^s_{f_1}(x_3)+(1-x_2-x_3)
\phi^t_{f_1}(x_3)]+\phi_{P}^T(x_2)
[(x_2+x_3-1)\phi_{f_1}^s(x_3)\right.\right. \right.\non &&
\left. \left.\left.
+(1-x_2+x_3)\phi_{f_1}^t(x_3)]\right)\right]
 \right\} \label{eq:pf1-nfa}\;,
\eeq
\item $(V-A)(V+A)$ operators:
\beq
M_{nfa}^{P1}&=& - \frac{32}{\sqrt{6}}\pi C_F m_{B}^2 \int_{0}^{1}d
x_{1}d x_{2}\,d x_{3}\,\int_{0}^{\infty} b_1d b_1 b_2d b_2\,
\phi_{B}(x_1,b_1)\left\{ \left[ r_0^P
x_2 \phi_{f_1}(x_3)(\phi_{P}^P(x_2)+\phi_{P}^T(x_2))
\right.\right.
 \non
 & & \left.\left.- r_{f_1} (1 -x_3)
 \phi_{P}^A(x_2)(\phi^s_{f_1}(x_3)- \phi^t_{f_1}(x_3))\right]
 E_{nfa}(t_e)h_{nfa}^e(x_1,x_2,x_3,b_1,b_2)+ h_{nfa}^f(x_1,x_2,x_3,b_1,b_2)
 \right. \non & & \left. \times
 \left[ r_0^P
(2-x_2)(\phi_{P}^P(x_2)+\phi_{P}^T(x_2))
 \phi_{f_1}(x_3)-r_{f_1} (1+x_3)
\phi_{P}^A(x_2)(\phi^s_{f_1}(x_3)-
 \phi^t_{f_1}(x_3))\right]E_{nfa}(t_f)
\right\} \label{eq:pf1-nfa-p1}\;,
\eeq
\item $(S-P)(S+P)$ operators:
\beq
M_{nfa}^{P2}&=& \frac{32}{\sqrt{6}}\pi C_F m_{B}^2
\int_{0}^{1}d x_{1}d x_{2}\,d x_{3}\,\int_{0}^{\infty} b_1d b_1 b_2d
b_2\,\phi_{B}(x_1,b_1)\left\{
\left[(1-x_3)\phi_{P}^A(x_2)\phi_{f_1}(x_3)+ r_0^P  r_{f_1}
\left(\phi_{P}^P(x_2)
\right.\right.\right.
\non & &
\left.\left.\left. \times [(x_2-x_3+3)
\phi^s_{f_1}(x_3)-
(1-x_2-x_3)\phi^t_{f_1}(x_3)]+\phi_{P}^T(x_2) [(1-x_2-x_3)\phi_{f_1}^s(x_3)+(1-x_2+x_3)\right.\right.\right.
\non && \left.\left.\left.
\times\phi_{f_1}^t(x_3)]
\right)\right] E_{nfa}(t_f)
h_{nfa}^f(x_1,x_2,x_3,b_1,b_2)- E_{nfa}(t_e)h_{nfa}^e(x_1,x_2,x_3,b_1,b_2)
\left[x_2 \phi_{P}^A(x_2)\phi_{f_1}(x_3)
\right.\right. \non && \left.\left. + r_0^P   r_{f_1}
\left( x_2(\phi_P^P(x_2)+\phi_P^T(x_2))(\phi_{f_1}^s(x_3)- \phi_{f_1}^t(x_3))
+(1-x_3)(\phi_P^P(x_2)- \phi_P^T(x_2))(\phi_{f_1}^s(x_3)+ \phi_{f_1}^t(x_3)\right)\right]
 \right\} \label{eq:pf1-nfa-p2}\;;
\eeq
\end{itemize}

For the last two diagrams in Fig.~\ref{fig:fig1}, i.e., the factorizable annihilation($fa$) diagrams
\ref{fig:fig1}(g) and \ref{fig:fig1}(h), we have
\begin{itemize}
\item $(V-A)(V-A)$ operators:
\beq
F_{fa}&=&  -8 \pi C_F m_{B}^2\int_0^1 d x_{2} dx_{3}\,
\int_{0}^{\infty} b_2 db_2 b_3 db_3\, \left\{ \left[ x_2
\phi_{P}^A(x_2) \phi_{f_1}(x_3)-2 r_0^P  r_{f_1}   \left((x_2 +
1)\phi^P_{P}(x_2)+(x_2-1)\phi^T_{P}(x_2)\right)\right.\right. \non & &
\left.\left.\times \phi_{f_1}^s(x_3)\right]
 h_{fa}(x_2,1-x_3,b_2,b_3)
E_{fa}(t_g)  - \left[(1-x_3)\phi_{P}^A(x_2) \phi_{f_1}(x_3)-2 r_0^P  r_{f_1}
\phi_{P}^P(x_2) \left((x_3-2)\phi^s_{f_1} (x_3)
\right.\right.\right.\non && \left.\left.\left.- x_3
\phi_{f_1}^t(x_3)\right) \right] E_{fa}(t_h)h_{fa}(1-x_3,x_2,b_3,b_2)
\right\}\label{eq:pf1-fa}\;,
\eeq
\item $(V-A)(V+A)$ operators:
\beq
 F_{fa}^{P1}&=& F_{fa}\;, \label{eq:pf1-fa-p1}
\eeq
\item $(S-P)(S+P)$ operators:
\beq
 F_{fa}^{P2}&=& -16 \pi C_F m_{B}^2  \int_0^1 d
x_{2} dx_{3}\, \int_{0}^{\infty} b_2 db_2 b_3 db_3\,\left\{ \left[2
r_{f_1} \phi_{P}^A(x_2) \phi^s_{f_1}(x_3)  +r_0^P
 x_2 (\phi_{P}^P(x_2)- \phi_{P}^T(x_2))\phi_{f_1}(x_3) \right]
\right. \non & &
\left. \times E_{fa}(t_g) h_{fa}(x_2,1-x_3,b_2,b_3)  + \left[ r_{f_1}
(1-x_3) \phi_{P}^A(x_2) (\phi_{f_1}^s(x_3)+\phi_{f_1}^t(x_3))+ 2
r_0^P \phi_P^P(x_2)\phi_{f_1}(x_3) \right]
 \right. \non && \left. \times
E_{fa}(t_h)h_{fa}(1-x_3,x_2,b_3,b_2) \right\} \label{eq:pf1-fa-p2}\;.
 \eeq
\end{itemize}

When we exchange $P$ and $f_1$ in Fig.~\ref{fig:fig1}, we can obtain the new
eight diagrams contributing to the considered $B \to f_1 P$ decays. The corresponding
factorization formulas can be easily obtained with the simple replacements in
Eqs.~(\ref{eq:pf1-fe})-(\ref{eq:pf1-fa-p2}) as follows,
\beq
f_{P} &\longleftrightarrow& f_{f_1}\;, \qquad
\phi_{P}^A \longleftrightarrow \phi_{f_1}\;, \qquad
\phi_{P}^P  \longleftrightarrow \phi_{f_1}^s\;, \qquad
\phi_{P}^T  \longleftrightarrow \phi_{f_1}^t\;, \qquad
r_0^P \longleftrightarrow r_{f_1} \;, \label{eq:replacements}
\eeq
where $F'$ and $M'$ will be used to denote the Feynman amplitudes from these new diagrams.
Note that, due to $\langle f_1 |S \pm P| 0 \rangle = 0$, the factorizable emission
amplitude $F_{fe}^{\prime P_2}$ is therefore absent naturally.

Before we write down the total decay amplitudes
for the $B \to f_1 P$ modes,
it is essential to give a brief discussion about the
$\eta-\eta'$ mixing and $f_{1}(1285)-f_{1}(1420)$ mixing, respectively.
The $\eta-\eta'$ mixing has been discussed in different bases:
quark-flavor basis~\cite{Feldmann:1998vh} and octet-singlet basis~\cite{Escribano:2005qq},
and the related parameters have been effectively constrained from
various experiments and theories(for a recently detailed overview, see~\cite{DiDonato:2011kr}
and references therein).
In the present work, we adopt the quark-flavor basis with the definitions~\cite{Feldmann:1998vh}
$\eta_q=(u\bar{u} + d\bar{d})/\sqrt{2}$ and $\eta_s=s\bar{s}$.
Then the physical states $\eta$ and $\eta^\prime$ can be described
as the mixtures of two quark-flavor $\eta_q$ and $\eta_s$ states with
a single mixing angle $\phi$,
\beq
\left( \begin{array}{c} \eta\\ \eta^\prime \\ \end{array} \right ) &=&
U(\phi) \left( \begin{array}{c}
 \eta_q\\ \eta_s \\ \end{array} \right ) =
  \left( \begin{array}{cc}
 \cos{\phi} & -\sin{\phi} \\
 \sin{\phi} & \cos\phi \end{array} \right )
 \left( \begin{array}{c}  \eta_q\\ \eta_s \\ \end{array} \right ).
 \label{eq:mix-eta-etap}
\eeq
It is assumed that the $\eta_q$ and $\eta_s$ states have the
same light-cone distribution amplitudes as that of the pion
but with different decay constants $f_{\eta_q}$ and $f_{\eta_s}$ and
different chiral enhancement factors $m_0^{\eta_q}$ and $m_0^{\eta_s}$.
The $f_{\eta_q}$, $f_{\eta_s}$ and $\phi$ in the quark-flavor basis
have been extracted from various related experiments~\cite{Feldmann:1998vh,Escribano:2005qq}:
\beq
f_{\eta_q} = (1.07\pm 0.02) f_\pi, \quad
f_{\eta_s} = (1.34 \pm 0.06) f_\pi, \quad \phi = 39.3^\circ \pm 1.0^\circ.
\label{eq:e}
\eeq
And the chiral enhancement factors are chosen as
\beq
  m_0^{\eta_q}
 & =& \frac{1}{2m_q}[m^2_\eta\cos^2\phi+
  m_{\eta'}^2\sin^2\phi-\frac{\sqrt 2f_{\eta_s}}{f_{\eta_q}}(m_{\eta'}^2-m_\eta^2)\cos\phi\sin\phi],\\
  m_0^{\eta_s}
  &=& \frac{1}{2m_s}[m^2_{\eta'}\cos^2\phi+
  m_{\eta}^2\sin^2\phi-\frac{f_{\eta_q}}{\sqrt
  2f_{\eta_s}}(m_{\eta'}^2-m_\eta^2)\cos\phi\sin\phi].
\eeq
with no isospin violation, i.e., the mass $m_q = m_u = m_d$.
It is worth mentioning that the effects of the possible gluonic
component of the $\eta^\prime$ meson will not be considered here, since it
is small in size~\cite{Liu:2005mm,Wang:2005bk,Charng:2006zj,Escribano:2007cd}.

Likewise, by considering both $f_1$ mesons as
the mixed quark flavor states,
then this $f_1(1285)-f_1(1420)$ mixing can also be described
as a $2\times 2$ rotation matrix with a single angle $\phi_{f_1}$
in the quark-flavor basis~\cite{Aaij:2013rja}, although there are also 
two mixing schemes for the $f_1(1285)-f_1(1420)$ mixing
system~\cite{Yang:2007zt,Cheng:2007mx,Yang:2010ah,Cheng:2011pb,Close:1997nm,Cheng:2013cwa}:
 \beq
\left(
\begin{array}{c} f_1(1285)\\ f_1(1420) \\ \end{array} \right ) &=&
  \left( \begin{array}{cc}
 \cos{\phi_{f_1}} & -\sin{\phi_{f_1}} \\
 \sin{\phi_{f_1}} & \cos{\phi_{f_1}} \end{array} \right )
 \left( \begin{array}{c}  f_{1q}\\ f_{1s} \\ \end{array} \right )\;.
 \label{eq:mix-f1-f1p}
 \eeq
As discussed in Ref.~\cite{Liu:2014doa},
since the axial-vector mesons have similar behavior to that
 of the vector ones, and the vector mesons $\rho$ and $\omega$ have the same distribution
amplitudes, except for the different decay constants $f_\rho$ and $f_\omega$,
we therefore assume that the $f_{1q}$ distribution amplitude 
is the same one as $a_1(1260)$ with decay constant 
$f_{f_{1q}}= 0.193^{+0.043}_{-0.038}$~GeV
~\cite{Verma:2011yw}.
For the $f_{1s}$ state, for the sake of simplicity,
we adopt the same distribution amplitude as flavor singlet $f_1$ state [not to be confused with
the abbreviation $f_1$ of $f_1(1285)$ and $f_1(1420)$ mesons]~\cite{Liu:2014doa}, but
with decay constant $f_{f_{1s}}= 0.230 \pm 0.009$~GeV~\cite{Verma:2011yw}.
For the masses of two $f_{1q}$ and $f_{1s}$ states, we adopt $m_{f_{1q}} \sim m_{f_1(1285)}$
and $m_{f_{1s}} \sim m_{f_1(1420)}$ for convenience.
Another more important factor is the value of the mixing angle $\phi_{f_1}$,
which is less constrained from the experiments currently.
Up to now, we just have some limited information on $\phi_{f_1}$ still
in controversy at the theoretical
and experimental aspects: (1) $(15^{+5}_{-10})^\circ$
provided by the Mark-II detector at SLAC from the ratio of $\frac{\Gamma(f_1(1285) \to
\gamma \gamma^*)}{\Gamma(f_1(1420) \to \gamma \gamma^*)}$~\cite{Gidal:1987bn};
(2) $(15.8^{+4.5}_{-4.6})^\circ$ extracted from the radiative $f_1(1285) \to \phi \gamma$ and
$\rho \gamma$ decays~\cite{Yang:2010ah} phenomenologically;
(3) $(27 \pm 2)^\circ$ from the updated LQCD calculations~\cite{Dudek:2013yja};
and (4) $(24.0^{+3.2}_{-2.7})^\circ$ measured first from the $B_{d/s} \to J/\psi f_1(1285)$
decays by the LHCb Collaboration~\cite{Aaij:2013rja} very recently. In view of the good consistency
for the values of $\phi_{f_1}$ between the latest measurements in $B$ meson
decays and the latest calculations in
LQCD, we will adopt
$\phi_{f_1} = 24^\circ$ as input in the numerical
evaluations.

Thus, by combining various contributions from different diagrams,
the total decay amplitudes for 20 charmless hadronic $B \to f_1 P$
decays in the pQCD approach can be written as
\begin{itemize}

\item[1.]{ $B^+ \to f_1(1285) (\pi^+,K^+)$ decays}
\beq
{\cal A}(B^+ \to f_1(1285) \pi^+) &=&
\biggl\{ [a_1] (f_\pi F_{fe}+ f_B F_{fa}+ f_B F'_{fa}) + [a_2] f_{f_{1q}} F'_{fe}
 + [C_1] (M_{nfe} + M_{nfa} + M'_{nfa}) \non &&  +[C_2] M'_{nfe}
\biggr\} \lambda_u^d \zeta_{f_{1q}} 
- \lambda_t^d \zeta_{f_{1q}}\biggl\{ [a_4 + a_{10}] (f_\pi F_{fe}+
f_B F_{fa} + f_B F'_{fa}) +
[a_6 + a_8]\non && \cdot (f_\pi F_{fe}^{P_2}+ f_B F_{fa}^{P_2}+ f_B F_{fa}^{\prime P_2}) + [C_3+ C_9] (M_{nfe}+ M_{nfa}
+ M'_{nfa})+ [C_5 + C_7] \non && \cdot (M_{nfe}^{P_1} + M_{nfa}^{P_1} +
M_{nfa}^{\prime P_1}) + [2a_3+ a_4 -2 a_5 -\frac{1}{2} (a_7 - a_9 + a_{10})] f_{f_{1q}} F'_{fe} + [C_3\non && + 2 C_4 - \frac{1}{2} (C_9 -C_{10})] M'_{nfe} +[C_5 - \frac{1}{2} C_7] M_{nfe}^{\prime P_1}
+[2 C_6 +\frac{1}{2} C_8] M_{nfe}^{\prime P_2}
\biggr\}   
-\lambda_t^d  \zeta_{f_{1s}} \non &&
\cdot  
\biggl\{ [a_3 -a_5 + \frac{1}{2} (a_7 - a_9)] f_{f_{1s}} F'_{fe} + [C_4 -\frac{1}{2} C_{10}] M'_{nfe}
+[C_6 - \frac{1}{2} C_8] M_{nfe}^{\prime P_2} \biggr\}
\label{eq:f1285pip};
\eeq
\beq
{\cal A}(B^+ \to f_1(1285) K^+) &=&
\lambda_u^s \biggl\{ [a_1] \biggl((f_K F_{fe}+ f_B F_{fa})  \zeta_{f_{1q}} 
+ f_B F'_{fa} \zeta_{f_{1s}} 
\biggr) + [a_2] f_{f_{1q}} F'_{fe}  \zeta_{f_{1q}} 
 + [C_1]
 \biggl(M'_{nfa} \zeta_{f_{1s}} 
 \non &&  +
 (M_{nfe} + M_{nfa})  \zeta_{f_{1q}} 
 \biggr) +[C_2] M'_{nfe}  \zeta_{f_{1q}} 
\biggr\}  - \lambda_t^s \biggl\{ [a_4 + a_{10}]
\biggl((f_K F_{fe}+
f_B F_{fa})  \zeta_{f_{1q}} 
\non &&
+ f_B F'_{fa} \zeta_{f_{1s}} 
\biggr) +
[a_6 + a_8]\biggl((f_K F_{fe}^{P_2}+ f_B F_{fa}^{P_2})  \zeta_{f_{1q}} 
+ f_B F_{fa}^{\prime P_2} \zeta_{f_{1s}} 
\biggr) + [C_3+ C_9] \non &&
\cdot \biggl(M'_{nfa} \zeta_{f_{1s}} 
+
(M_{nfe}+ M_{nfa})  \zeta_{f_{1q}} 
\biggr)+ [C_5
+ C_7] \biggl((M_{nfe}^{P_1} + M_{nfa}^{P_1})  \zeta_{f_{1q}} 
+
M_{nfa}^{\prime P_1} \zeta_{f_{1s}} 
\biggr) \non &&
+ \biggl([2a_3 -2 a_5 -\frac{1}{2} (a_7 - a_9)] f_{f_{1q}}  F'_{fe}
+ [  2 C_4 + \frac{1}{2} C_{10}]M'_{nfe}
+[2 C_6 +\frac{1}{2} C_8] M_{nfe}^{\prime P_2}\biggr)
\zeta_{f_{1q}} 
 \non && +\biggl( [a_3 +a_4 -a_5 + \frac{1}{2} (a_7 - a_9
 - a_{10})] f_{f_{1s}} F'_{fe} + [C_3+C_4 -\frac{1}{2}(C_9+ C_{10})] M'_{nfe}
\non &&  +[C_5 - \frac{1}{2} C_7] M_{nfe}^{\prime P_1}
+[C_6 - \frac{1}{2} C_8]  M_{nfe}^{\prime P_2}\biggr)  \zeta_{f_{1s}} 
\biggr\}
\label{eq:f1285kp};
\eeq

\item[2.]{$B_d^0 \to f_1(1285) (\pi^0, K^0, \eta, \eta')$ decays}
\beq
\sqrt{2}{\cal A}(B_d^0 \to f_1(1285) \pi^0) &=&
\biggl\{ [a_2] (f_\pi F_{fe}+ f_B F_{fa}+ f_B F'_{fa} -f_{f_{1q}} F'_{fe})
 + [C_2] (M_{nfe} + M_{nfa} + M'_{nfa} -  M'_{nfe})
\biggr\} \non && \cdot\lambda_u^d \zeta_{f_{1q}} 
- \lambda_t^d \zeta_{f_{1q}} 
\biggl\{ [-a_4 -\frac{1}{2}(3 a_7- 3a_{9}- a_{10})] f_\pi F_{fe}+[-a_4 +\frac{1}{2}(3 a_7+ 3a_{9}+ a_{10})]\non && \cdot (f_B F_{fa} + f_B F'_{fa})- [2 a_3 +a_4- 2a_5 -\frac{1}{2}(a_7- a_{9}+ a_{10})]f_{f_{1q}} F'_{fe}
-[a_6 - \frac{1}{2} a_8] (f_\pi \non && \cdot
F_{fe}^{P_2}+ f_B F_{fa}^{P_2}+ f_B F_{fa}^{\prime P_2}) + [-C_3+ \frac{1}{2}(C_9+3 C_{10})] (M_{nfe}+ M_{nfa}
+ M'_{nfa})
+[\frac{3}{2} C_8] \non &&\cdot (M_{nfe}^{P_2}+ M_{nfa}^{P_2}
+ M_{nfa}^{\prime P_2})
-[C_5 -\frac{1}{2} C_7]  (M_{nfe}^{P_1} + M_{nfa}^{P_1} +
M_{nfa}^{\prime P_1}+M_{nfe}^{\prime P_1})  -[C_3 + 2 C_4 \non &&- \frac{1}{2} (C_9 -C_{10})] M'_{nfe} -[2 C_6 + \frac{1}{2} C_8] M_{nfe}^{\prime P_2}
\biggr\}
-\lambda_t^d \biggl\{ -[a_3 -a_5 + \frac{1}{2} (a_7 - a_9)] f_{f_{1s}} F'_{fe} \non && - [C_4 -\frac{1}{2} C_{10}] M'_{nfe}
-[C_6 - \frac{1}{2} C_8] M_{nfe}^{\prime P_2} \biggr\}  \zeta_{f_{1s}} 
\label{eq:f1285pi-d};
\eeq
\beq
{\cal A}(B_d^0 \to f_1(1285) K^0) &=&
\lambda_u^s \biggl\{
[a_2] f_{f_{1q}} F'_{fe}
 +[C_2] M'_{nfe}
\biggr\} \zeta_{f_{1q}} 
- \lambda_t^s \biggl\{ [a_4 -\frac{1}{2} a_{10}] \biggl((f_K F_{fe}+
f_B F_{fa}) \zeta_{f_{1q}} 
+\zeta_{f_{1s}} 
\non && \cdot
 f_B F'_{fa}
\biggr) +
[a_6 -\frac{1}{2} a_8] \biggl((f_K F_{fe}^{P_2}+ f_B F_{fa}^{P_2}) \zeta_{f_{1q}}
+ f_B F_{fa}^{\prime P_2}\zeta_{f_{1s}} 
\biggr) + [C_3-\frac{1}{2} C_9]
\non && \cdot \biggl((M_{nfe}+ M_{nfa})
\zeta_{f_{1q}} 
+ M'_{nfa}\zeta_{f_{1s}} 
\biggr)
+ [C_5 -\frac{1}{2} C_7]  \biggl((M_{nfe}^{P_1}
 + M_{nfa}^{P_1}) \zeta_{f_{1q}} 
+ M_{nfa}^{\prime P_1}\zeta_{f_{1s}} 
\biggr)
\non && + \biggl([2a_3 -2 a_5 -\frac{1}{2} (a_7 - a_9)] f_{f_{1q}} F'_{fe}
+ [  2 C_4 + \frac{1}{2} C_{10}] M'_{nfe}
+[2 C_6 +\frac{1}{2} C_8] M_{nfe}^{\prime P_2}\biggr) \zeta_{f_{1q}}
\non && +\biggl( [a_3 +a_4 -a_5 + \frac{1}{2} (a_7 - a_9 - a_{10})]
f_{f_{1s}} F'_{fe} + [C_3+C_4 -\frac{1}{2}(C_9+ C_{10})] M'_{nfe}\non &&
+[C_5 - \frac{1}{2} C_7] M_{nfe}^{\prime P_1}
+[C_6 - \frac{1}{2} C_8]  M_{nfe}^{\prime P_2}\biggr) \zeta_{f_{1s}} 
\biggr\}
\label{eq:f1285k0};
\eeq
\beq
{\cal A}(B_d^0 \to f_1(1285) \eta) &=&
\lambda_u^d \biggl\{ [a_2] (f_{\eta_q} F_{fe}+ f_B F_{fa}+ f_B F'_{fa} +f_{f_{1q}} F'_{fe})
 + [C_2] (M_{nfe} + M_{nfa} + M'_{nfa} +  M'_{nfe})
\biggr\} \non && \cdot\zeta_{f_{1q}} 
\cdot \zeta_{\eta_q} 
- \lambda_t^d \biggl\{[2 a_3 +a_4- 2a_5 -\frac{1}{2}(a_7- a_{9}+ a_{10})](f_{\eta_q} F_{fe} +f_{f_{1q}} F'_{fe})+ [2 a_3+a_4 \non && + 2a_5 +\frac{1}{2}(a_7+ a_{9}- a_{10})] (f_B F_{fa} + f_B F'_{fa})
+[a_6 - \frac{1}{2} a_8] (f_{\eta_q} F_{fe}^{P_2}+ f_B F_{fa}^{P_2}+ f_B F_{fa}^{\prime P_2})
\non &&  +[C_3 + 2 C_4 - \frac{1}{2} (C_9 -C_{10})] (M_{nfe}+ M'_{nfe} + M_{nfa}
+ M'_{nfa})
+[C_5 -\frac{1}{2} C_7] (M_{nfe}^{P_1}\non &&   + M_{nfa}^{P_1} +
M_{nfa}^{\prime P_1}+M_{nfe}^{\prime P_1})+[2 C_6 + \frac{1}{2} C_8] (M_{nfe}^{P_2}+ M_{nfe}^{\prime P_2}+ M_{nfa}^{P_2}
+ M_{nfa}^{\prime P_2})
\biggr\}\cdot  \zeta_{f_{1q}} 
\cdot \zeta_{\eta_q} 
\non && -\lambda_t^d \biggl\{ [a_3 -a_5 + \frac{1}{2} (a_7 - a_9)] \biggl(f_{\eta_s} F_{fe}
\zeta_{\eta_s} 
\cdot \zeta_{f_{1q}} 
+ f_{f_{1s}} F'_{fe}
\cdot\zeta_{f_{1s}} 
\cdot \zeta_{\eta_q} 
\biggr) +[a_3 +a_5- \frac{1}{2} \non && \cdot (a_7 + a_9)] (f_B F_{fa}+ f_B F'_{fa})
\zeta_{\eta_s} 
\cdot \zeta_{f_{1s}} 
+[C_4 -\frac{1}{2} C_{10}] \biggl(M_{nfe}\zeta_{\eta_s} 
\cdot \zeta_{f_{1q}} 
+M'_{nfe}\zeta_{f_{1s}} 
\cdot \zeta_{\eta_q} 
\non &&
+(M_{nfa}+M'_{nfa})
\zeta_{\eta_s} 
\zeta_{f_{1s}} 
\biggr)
+[C_6 - \frac{1}{2} C_8] \biggl(M_{nfe}^{P_2}\zeta_{\eta_s} 
\cdot
\zeta_{f_{1q}} 
+M_{nfe}^{\prime P_2}\zeta_{f_{1s}} 
\cdot \zeta_{\eta_q} 
+
(M_{nfa}^{P_2}\non &&  +M_{nfa}^{\prime P_2})
\zeta_{\eta_s} 
\cdot
\zeta_{f_{1s}} 
\biggr) \biggr\}
\label{eq:f1285eta-d};
\eeq
\beq
{\cal A}(B_d^0 \to f_1(1285) \eta') &=&
 \biggl\{ [a_2] (f_{\eta_q} F_{fe}+ f_B F_{fa}+ f_B F'_{fa} +f_{f_{1q}} F'_{fe})
 + [C_2] (M_{nfe} + M_{nfa} + M'_{nfa} +  M'_{nfe})
\biggr\} \non && \cdot \lambda_u^d\cdot\zeta_{f_{1q}} 
\cdot \zeta'_{\eta_q} 
- \lambda_t^d \biggl\{[2 a_3 +a_4- 2a_5 -\frac{1}{2}(a_7- a_{9}+ a_{10})](f_{\eta_q} F_{fe} +f_{f_{1q}} F'_{fe})+ [2 a_3\non && +a_4+ 2a_5 +\frac{1}{2}(a_7+ a_{9}- a_{10})] (f_B F_{fa} + f_B F'_{fa})
+[a_6 - \frac{1}{2} a_8] (f_{\eta_q} F_{fe}^{P_2}+ f_B F_{fa}^{P_2}\non && + f_B  F_{fa}^{\prime P_2})+[C_3 + 2 C_4 - \frac{1}{2} (C_9 -C_{10})] (M_{nfe}+ M'_{nfe} + M_{nfa}
+ M'_{nfa})
+[C_5 -\frac{1}{2} C_7] \non && \cdot (M_{nfe}^{P_1} + M_{nfa}^{P_1} +
M_{nfa}^{\prime P_1}+M_{nfe}^{\prime P_1})+[2 C_6 + \frac{1}{2} C_8] (M_{nfe}^{P_2}+ M_{nfe}^{\prime P_2}+ M_{nfa}^{P_2}
+ M_{nfa}^{\prime P_2})
\biggr\} \non && \cdot \zeta_{f_{1q}} 
\cdot \zeta'_{\eta_q} 
-\lambda_t^d \biggl\{ [a_3 -a_5 + \frac{1}{2} (a_7 - a_9)] \biggl(f_{\eta_s} F_{fe}
\zeta'_{\eta_s} 
\cdot \zeta_{f_{1q}}
+ f_{f_{1s}} F'_{fe} \cdot\zeta_{f_{1s}} 
\cdot \zeta'_{\eta_q} 
\biggr) +[a_3  \non && +a_5- \frac{1}{2} (a_7 + a_9)]
(f_B F_{fa}+ f_B F'_{fa})
\zeta'_{\eta_s} 
\cdot \zeta_{f_{1s}} 
+[C_4 -\frac{1}{2} C_{10}] \biggl(M_{nfe}
\zeta'_{\eta_s} 
\cdot
\zeta_{f_{1q}} 
+M'_{nfe}
\non && \cdot
\zeta_{f_{1s}} 
\cdot \zeta'_{\eta_q} 
+(M_{nfa}+M'_{nfa})
\zeta'_{\eta_s} 
\cdot \zeta_{f_{1s}} 
\biggr)
+[C_6 - \frac{1}{2} C_8] \biggl(M_{nfe}^{P_2}\zeta'_{\eta_s} 
\cdot \zeta_{f_{1q}}
+M_{nfe}^{\prime P_2}\zeta_{f_{1s}} 
\cdot\zeta'_{\eta_q} 
\non &&  +
(M_{nfa}^{P_2} +M_{nfa}^{\prime P_2})
\zeta'_{\eta_s} 
\cdot \zeta_{f_{1s}} 
\biggr) \biggr\}
\label{eq:f1285etap-d};
\eeq

\item[3.]{$B_s^0 \to f_1(1285) (\pi^0, \bar{K}^0, \eta, \eta')$ decays}
\beq
\sqrt{2}{\cal A}(B_s^0 \to f_1(1285) \pi^0) &=&
\biggl\{ [a_2] \biggr(f_\pi F_{fe}\zeta_{f_{1s}} 
+ (f_B F_{fa}+ f_B F'_{fa})\zeta_{f_{1q}} 
\biggl)
 + [C_2] \biggr(M_{nfe} \zeta_{f_{1s}} 
 + (M_{nfa} + M'_{nfa})
 \non &&\cdot \zeta_{f_{1q}} 
 \biggl)
\biggr\}\lambda_u^s    - \lambda_t^s \biggl\{ \frac{3}{2}[a_{9}- a_{7}] f_\pi F_{fe}
\zeta_{f_{1s}} 
+\frac{3}{2}[ a_7+ a_{9}](f_B F_{fa} + f_B F'_{fa})
\zeta_{f_{1q}} 
+ \frac{3}{2} C_{10} \non &&  \cdot  \biggr(M_{nfe}\zeta_{f_{1s}} 
+ (M_{nfa}
+ M'_{nfa})\zeta_{f_{1q}} 
\biggl)
+\frac{3}{2} [C_8] \biggr(M_{nfe}^{P_2}\zeta_{f_{1s}} 
+ (M_{nfa}^{P_2}
+ M_{nfa}^{\prime P_2})\zeta_{f_{1q}} 
\biggl) \biggr\}
\label{eq:f1285pi-s};
\eeq
\beq
{\cal A}(B_s^0 \to f_1(1285) \bar K^0) &=&
\lambda_u^d \biggl\{
[a_2] f_{f_{1q}} F'_{fe}
 +[C_2] M'_{nfe}
\biggr\} \zeta_{f_{1q}} 
- \lambda_t^d \biggl\{ [a_4 -\frac{1}{2} a_{10}] \biggl( (f_K F_{fe}+
f_B F_{fa}) \zeta_{f_{1s}} 
+ f_B F'_{fa}\non &&
\cdot \zeta_{f_{1q}} 
\biggr)+
[a_6 -\frac{1}{2} a_8] \biggl((f_K F_{fe}^{P_2}+ f_B F_{fa}^{P_2})
\zeta_{f_{1s}} 
+ f_B F_{fa}^{\prime P_2}\zeta_{f_{1q}} 
\biggr) + [C_3-\frac{1}{2} C_9] \biggl(
 M'_{nfa}\zeta_{f_{1q}} 
\non &&
+(M_{nfe}+ M_{nfa})
\zeta_{f_{1s}} 
\biggr)
+ [C_5 -\frac{1}{2} C_7]  \biggl((M_{nfe}^{P_1} + M_{nfa}^{P_1})
\zeta_{f_{1s}} 
+
M_{nfa}^{\prime P_1}\zeta_{f_{1q}} 
\biggr)
+ \biggl(  [2a_3 \non &&
+ a_4
-2 a_5 -\frac{1}{2} (a_7 - a_9+ a_{10})] f_{f_{1q}} F'_{fe}
+ [ C_3 + 2 C_4 - \frac{1}{2}(C_9 - C_{10})] M'_{nfe}
+[C_5 \non && - \frac{1}{2} C_7] M_{nfe}^{\prime P_1}
+[2 C_6 +\frac{1}{2} C_8] M_{nfe}^{\prime P_2}\biggr)
\zeta_{f_{1q}} 
+ \biggl( [a_3 -a_5 + \frac{1}{2} (a_7 - a_9 )]f_{f_{1s}} F'_{fe}
+ [C_4 \non &&  -\frac{1}{2} C_{10}] M'_{nfe}
+[C_6 - \frac{1}{2} C_8]  M_{nfe}^{\prime P_2}\biggr)
\zeta_{f_{1s}} 
\biggr\}
\label{eq:f1285k0b};
\eeq
\beq
{\cal A}(B_s^0 \to f_1(1285) \eta) &=&
\lambda_u^s \biggl\{
\zeta_{f_{1s}} 
\cdot \zeta_{\eta_q} 
([a_2] f_{\eta_q} F_{fe}+[C_2] M_{nfe})
+\zeta_{\eta_s} 
\cdot \zeta_{f_{1q}} 
([a_2] f_{f_{1s}} F'_{fe}+ [C_2]
M'_{nfe})+ \zeta_{\eta_q} 
\cdot
\zeta_{f_{1q}} 
\non && \cdot \biggl( [a_2] (f_B F_{fa}+ f_B F'_{fa})
 + [C_2] ( M_{nfa} + M'_{nfa} )\biggr)
\biggr\}
- \lambda_t^s \biggl\{ \zeta_{\eta_s} 
\cdot \zeta_{f_{1s}} 
\biggl(
(a_3+a_4-a_5+\frac{1}{2}
\non && \cdot (a_7-a_9-a_{10})) (f_{\eta_s} F_{fe} + f_{1s} F'_{fe} )
+(a_6-\frac{1}{2}a_8) (f_{\eta_s} F_{fe}^{P_2}
+f_B F_{fa}^{P_2} + f_B F_{fa}^{\prime P_2} )
+(C_3\non &&
+C_4-\frac{1}{2}(C_9+C_{10}))(M_{nfe} + M'_{nfe}+M_{nfa} + M'_{nfa})
+(C_5 - \frac{1}{2} C_7) (M_{nfe}^{P_1} + M_{nfe}^{\prime P_1}\non &&
+M_{nfa}^{P_1} + M_{nfa}^{\prime P_1})
+(C_6 - \frac{1}{2} C_8) (M_{nfe}^{P_2} + M_{nfe}^{\prime P_2}+M_{nfa}^{P_2} + M_{nfa}^{\prime P_2})
+(a_3+a_4+a_5
\non &&-\frac{1}{2}(a_7+a_9+a_{10})) (f_B F_{fa} +f_B F'_{fa}) \biggr)
+\zeta_{\eta_q} 
\cdot
\zeta_{f_{1q}} 
\biggl( (2C_4 + \frac{1}{2} C_{10})
(M_{nfa} + M'_{nfa})\non &&
+ (2C_6 + \frac{1}{2} C_{8}) (M_{nfa}^{P_2} + M_{nfa}^{\prime P_2})
+ (2a_3+2a_5+\frac{1}{2}(a_7+a_9)) (f_B F_{fa}
+ f_B F'_{fa}) \biggr)
+ \zeta_{f_{1s}} 
\non &&\cdot \zeta_{\eta_q} 
\biggl( (2a_3 -2a_5 - \frac{1}{2}(a_7 - a_9)) f_{\eta_q} F_{fe}
+(2C_4 + \frac{1}{2} C_{10}) M_{nfe}
+ (2C_6 + \frac{1}{2} C_8) M_{nfe}^{P_2} \biggr)
\non &&+  \zeta_{\eta_s} 
\cdot \zeta_{f_{1q}} 
\biggl( (2a_3 -2a_5 - \frac{1}{2}(a_7 - a_9)) f_{1s} F'_{fe}
+(2C_4
+ \frac{1}{2} C_{10}) M'_{nfe} + (2C_6 + \frac{1}{2} C_8) M_{nfe}^{\prime P_2} \biggr) \biggr\}
\label{eq:f1285eta-s};
\eeq
\beq
{\cal A}(B_s^0 \to f_1(1285) \eta') &=&
\lambda_u^s \biggl\{
\zeta_{f_{1s}} 
\cdot \zeta'_{\eta_q} 
([a_2] f_{\eta_q} F_{fe}+[C_2] M_{nfe})
+\zeta'_{\eta_s} 
\cdot \zeta_{f_{1q}} 
([a_2] f_{f_{1s}} F'_{fe}+ [C_2]
M'_{nfe})+ \zeta'_{\eta_q} 
\cdot \zeta_{f_{1q}} 
\non &&
\cdot \biggl( [a_2] (f_B F_{fa}+ f_B F'_{fa})
 + [C_2] ( M_{nfa} + M'_{nfa} )\biggr)
\biggr\}
- \lambda_t^s \biggl\{ \zeta'_{\eta_s} 
\cdot \zeta_{f_{1s}} 
\biggl(
(a_3+a_4-a_5+\frac{1}{2}
\non && \cdot
(a_7-a_9-a_{10})) (f_{\eta_s} F_{fe} + f_{1s} F'_{fe} )
+(a_6-\frac{1}{2}a_8) (f_{\eta_s} F_{fe}^{P_2}
+f_B F_{fa}^{P_2} + f_B F_{fa}^{\prime P_2} )
+(C_3\non &&
+C_4-\frac{1}{2}(C_9+C_{10}))(M_{nfe} + M'_{nfe}+M_{nfa} + M'_{nfa})
+(C_5
- \frac{1}{2} C_7) (M_{nfe}^{P_1} + M_{nfe}^{\prime P_1}\non &&
+M_{nfa}^{P_1} + M_{nfa}^{\prime P_1})
+(C_6 - \frac{1}{2} C_8) (M_{nfe}^{P_2} + M_{nfe}^{\prime P_2}+M_{nfa}^{P_2} + M_{nfa}^{\prime P_2})
+(a_3+a_4+a_5-\frac{1}{2}
\non &&\cdot (a_7+a_9+a_{10})) (f_B F_{fa} +f_B F'_{fa}) \biggr)
+\zeta'_{\eta_q} 
\cdot
\zeta_{f_{1q}} 
\biggl( (2C_4 + \frac{1}{2} C_{10})
(M_{nfa} + M'_{nfa})
+ (2C_6  \non && + \frac{1}{2} C_{8})(M_{nfa}^{P_2} + M_{nfa}^{\prime P_2})
+ (2a_3+2a_5+\frac{1}{2}(a_7+a_9)) (f_B F_{fa}
+ f_B F'_{fa}) \biggr)
+ \zeta_{f_{1s}} 
\cdot \zeta'_{\eta_q} 
\biggl( (2a_3\non && -2a_5 - \frac{1}{2}(a_7 - a_9)) f_{\eta_q} F_{fe}
+(2C_4 + \frac{1}{2} C_{10}) M_{nfe}
+ (2C_6 + \frac{1}{2} C_8) M_{nfe}^{P_2} \biggr)
+  \zeta'_{\eta_s} 
\cdot \zeta_{f_{1q}} 
\biggl( (2a_3 \non && -2a_5 - \frac{1}{2}(a_7 - a_9)) f_{1s} F'_{fe}
+(2C_4
+ \frac{1}{2} C_{10}) M'_{nfe} + (2C_6 + \frac{1}{2} C_8) M_{nfe}^{\prime P_2} \biggr) \biggr\}
\label{eq:f1285etap-s};
\eeq

\end{itemize}
where $\lambda_u^{d(s)}=V^*_{ub} V_{ud(s)}$ and  $\lambda_t^{d(s)}=
V^*_{tb} V_{td(s)}$, $\zeta_{f_{1q}} = \cos\phi_{f_1}/\sqrt{2}$ and
$\zeta_{f_{1s}}= -\sin\phi_{f_1}$, $\zeta_{\eta_q} = \cos\phi/\sqrt{2}$
and $\zeta_{\eta_s}= -\sin\phi$, and $\zeta'_{\eta_q}= \sin\phi/\sqrt{2}$
and $\zeta'_{\eta_s}=\cos\phi$.
When we make the replacements with  $\zeta_{f_{1q}} \to \zeta'_{f_{1q}} \sim \sin\phi_{f_1}
/\sqrt{2}$ and $\zeta_{f_{1s}} \to \zeta'_{f_{1s}} \sim \cos\phi_{f_1}$ in the above equations, i.e., Eqs.~(\ref{eq:f1285pip})-(\ref{eq:f1285etap-s}), the decay amplitudes of $B \to f_1(1420) P$ modes
will be easily obtained.

\section{Numerical Results and Discussions} \label{sec:randd}

In this section, we will present the theoretical predictions on the
\cp-averaged branching ratios and \cp-violating asymmetries
for the considered 20 $B \to f_1 P$ decay modes in the pQCD approach.
In numerical calculations, central values of the input parameters will be
used implicitly unless otherwise stated. The relevant QCD scale~({\rm GeV}), masses~({\rm GeV}),
and $B$ meson lifetime({\rm ps}) are the following ~\cite{Keum:2000ph,Lu:2000em,Yang:2007zt,Beringer:1900zz}:
\beq
 \Lambda_{\overline{\mathrm{MS}}}^{(f=4)} &=& 0.250\; , \quad m_W = 80.41\;,
 \quad  m_{B}= 5.28\;, \quad  m_{B_s}= 5.37\;, \quad  m_b = 4.8 \;; \non
  f_{\pi}&=& 0.13\;, \quad f_{K} = 0.16\;, \quad m_{f_1(1285)}= 1.2812 \;,
\quad m_{f_1(1420)}= 1.4264\;; \non
  m_0^{\pi}&=& 1.4\;, \quad m_0^{K} = 1.6\;, \quad m_0^{\eta_q}= 1.08 \;,
\quad m_0^{\eta_s}= 1.92\;, \quad \phi_{f_1} = 24.0^\circ\;; \non
  \tau_{B_u} &=& 1.641\;,  \quad \tau_{B_d}= 1.519\;,
   \quad  \tau_{B_s}= 1.497\;.
\label{eq:mass}
\eeq

For the CKM matrix elements, we adopt the Wolfenstein
parametrization and the updated parameters $A=0.811$,
 $\lambda=0.22535$, $\bar{\rho}=0.131^{+0.026}_{-0.013}$, and $\bar{\eta}=0.345^{+0.013}_{-0.014}$~\cite{Beringer:1900zz}.

\subsection{\cp-averaged branching ratios of $B \to f_1 P$ decays in the pQCD approach}

For the considered $B \to  f_1  P$ decays, the decay rate can be written as
\beq
\Gamma =\frac{G_{F}^{2}m^{3}_{B}}{32 \pi  } (1-  r_{f_1}^2) |{\cal A}(B
\to f_1 P)|^2\;,\label{eq:dr-f1p}
\eeq
where the corresponding decay amplitudes ${\cal A}$ have been
given explicitly in Eqs.~(\ref{eq:f1285pip})$\sim$(\ref{eq:f1285etap-s}).
Using the decay amplitudes obtained in the last section, 
it is straightforward to calculate the \cp-averaged
branching ratios with uncertainties
for the considered decay modes in the pQCD approach.
The pQCD predictions for the \cp-averaged
branching ratios of the considered $B \to f_1 P$ decays
have been collected in Tables~\ref{tab:BR-Bu} and
\ref{tab:BR-Bds}. Based on these numerical results, some phenomenological
discussions are given in order:
\begin{table}[hbt]
\caption{ The \cp-averaged branching ratios for $B^+ \to
f_1 (\pi^+, K^+)$ decays in the pQCD approach.}
\label{tab:BR-Bu}
 \begin{center}\vspace{-0.3cm}{
\begin{tabular}[t]{c|c}
\hline  \hline
  Channels   &   \cp-averaged branching ratios \\
   \hline
 $B^+ \to f_1(1285) \pi^+$
&$4.0^{+1.1}_{-0.8} (\omega_b)
^{+1.9}_{-1.4} (f_{f_1})
^{+2.2}_{-1.7}(a_i^M)
^{+0.2}_{-0.2} (\phi_{f_1})
^{+0.1}_{-0.1} (a_t)
\times 10^{-6}$
 \\
 $B^+ \to f_1(1420) \pi^+$
&$7.4^{+2.0}_{-1.5} (\omega_b)
^{+3.6}_{-2.6} (f_{f_1})
^{+4.1}_{-3.2}(a_i^M)
^{+1.9}_{-1.5} (\phi_{f_1})
^{+0.2}_{-0.2} (a_t)
\times 10^{-7}$
 \\
 $B^+ \to f_1(1285) K^+$
&$1.6^{+0.4}_{-0.3} (\omega_b)
^{+1.2}_{-0.8} (f_{f_1})
^{+1.8}_{-1.1}(a_i^M)
^{+0.2}_{-0.3} (\phi_{f_1})
^{+0.1}_{-0.1} (a_t)
\times 10^{-6}$
 \\
 $B^+ \to f_1(1420) K^+$
&$5.1^{+1.0}_{-0.8} (\omega_b)
^{+0.9}_{-0.7} (f_{f_1})
^{+1.4}_{-1.2}(a_i^M)
^{+0.3}_{-0.3} (\phi_{f_1})
^{+0.7}_{-0.6} (a_t)
\times 10^{-6}$
 \\
 \hline \hline
\end{tabular}}
\end{center}
\end{table}

\begin{enumerate}

\item[(1)]
The theoretical errors of these predictions in the pQCD approach
are induced mainly by the uncertainties
of the shape parameters $\omega_b = 0.40 \pm 0.04\ (\omega_b = 0.50 \pm 0.05)$~GeV
for the $B_{u,d}\ (B_s)$ meson wave function, of the combined $f_{f_1}$ from
the axial-vector $f_{1q(s)}$ state decay constant $f_{f_{1q}}=0.193^{+0.043}_{-0.038} 
(f_{f_{1s}}=0.230 \pm 0.009)$~GeV,
of the combined Gegenbauer moments $a_i^M$ from
$a_i^{\parallel,\perp}\ (i=1,2)$
for the axial-vector $f_{1q(s)}$ states in the longitudinal polarization
and $a_{(1)2}^P$ for the pseudoscalar $P$ meson,
and of the mixing angle $\phi_{f_1}= (24.0^{+3.2}_{-2.7})^\circ$, respectively.
Note that very small effects induced by the variation of
the CKM parameters appear in the \cp-averaged branching ratios of
these considered $B \to f_1 P$ decays
and thus have been safely neglected. Furthermore,
we also investigate the higher order contributions simply through
exploring the variation of the hard scale $t_{\rm max}$, i.e., from $0.8t$ to $1.2t$
(not changing $1/b_i, i= 1,2,3$), in the hard kernel,
which have been counted into one of the sources of theoretical uncertainties.
One can clearly observe that some penguin-dominated decays such as
$B^+ \to f_1(1420) K^+$, $B_d^0 \to f_1(1420) K^0$, $B_s^0 \to f_1 \bar K^0$,
and $B_{d/s}^0 \to f_1 \eta^{(\prime)}$ channels get large higher order
corrections around $15\% \sim 40\%$ to the \cp-averaged branching ratios
as presented in the Tables~\ref{tab:BR-Bu}$\sim$\ref{tab:BR-Bds}.

\item[(2)]
The considered $B \to f_1 P$ decays can be classified into two kinds of transitions, i.e.,
$b \to d(\Delta S =0)$ and $b \to s (\Delta S =1)$, respectively.
The former transition includes ten $B_{u,d} \to
f_1 (\pi, \eta, \eta')$ and $B_s \to f_1 \bar K^0$ modes, while the latter transition
contains the other ten $B_{u,d} \to f_1 K$ and $B_s \to f_1 (\pi^0, \eta, \eta')$ channels.
\begin{itemize}

\item[(a)]
For $\Delta S =0$ decays, it is found that most of the branching ratios are in
the order of $10^{-8} \sim 10^{-7}$ in the pQCD approach,
except for the $B^+ \to f_1 \pi^+$ modes
with the decay rates as
\beq
Br(B^+ \to f_1(1285) \pi^+) &\approx& 4.0^{+3.1}_{-2.4} \times 10^{-6} \;,
\qquad
Br(B^+ \to f_1(1420) \pi^+) \approx   0.7^{+0.6}_{-0.5} \times 10^{-6} \;;
\eeq
which are around ${\cal O}(10^{-6})$ within large errors, where various errors
as specified previously have been added in quadrature. It is noted that these two $B^+
\to f_1 \pi^+$ decays are dominated by the color-allowed tree amplitudes,
while the other eight $B_{d/s}^0 \to f_1 (\pi^0,\eta,\eta')/\bar K^0$ processes are
basically penguin dominant with color-suppressed tree contributions.
In particular, for the $B_s^0 \to f_1 \bar K^0$ channels, the tree
pollution is so tiny that it can be neglected safely for the predictions of the \cp-averaged
branching ratios.

\item[(b)]
For $\Delta S = 1$ decays, contrary to the $\Delta S =0$ ones,
it is observed that most of the branching ratios are in the order of $10^{-6} \sim 10^{-5}$
in the pQCD approach, apart from the
$B_s^0 \to f_1 \pi^0$ channels
with the decay rates as
\beq
Br(B_s^0 \to f_1(1285) \pi^0) &\approx& 2.7^{+2.0}_{-1.6} \times 10^{-8} \;,
\qquad
Br(B_s^0 \to f_1(1420) \pi^0) \approx   1.4^{+1.0}_{-0.7} \times 10^{-7} \;;
\eeq
in which the theoretical errors from the input parameters have also been added in
quadrature. In contrast to the above case, it is worthwhile to stress that all of the
$b \to s$ transition processes are determined by the penguin contributions dramatically
just with generally very small tree contaminations.

\end{itemize}
The relation of the \cp-averaged branching ratios between
these two $\Delta S =0$ and
$\Delta S =1$ transitions can be understood naively through the involved CKM hierarchy~\cite{Agashe:2014kda}, apart
from the the interferences between $f_{1q} P$ and $f_{1s} P$ states, $|\lambda_u^d|:|\lambda_u^s|:|\lambda_t^d|:|\lambda_t^s| \sim 0.09:0.02:0.22:1$, which means
that when the decays are dominated by the penguin contributions, then we must observe at least
one order difference as roughly anticipated because of the value around $21$ of
 $|\lambda_t^s/\lambda_t^d|^2$.
It is known that the $B_d^0 \to K^+ K^-$ with decay rate $1.3 \pm 0.5 \times 10^{-7}$ and
 the $B_s^0 \to \pi^+ \pi^-$ with branching ratio $7.6 \pm 1.9 \times 10^{-7}$
 have been detected by the experiments~\cite{Agashe:2014kda}.
Therefore, the decay modes with the branching ratios in the order of $10^{-6}$ and larger
are generally expected to be accessed more easily at the running LHCb and forthcoming 
Belle II 
experiments in the near future.

\begin{table}[b]
\caption{ Same as Table~\ref{tab:BR-Bu} but for $B_{d/s}^0 \to
f_1 (\pi^0, K^0, \eta, \eta')$ decays.
}
\label{tab:BR-Bds}
 \begin{center}\vspace{-0.3cm}{
\begin{tabular}[t]{c|c|c|c}
\hline  \hline
  Channels   &   \cp-averaged branching ratios
& Channels   &   \cp-averaged branching ratios \\
   \hline
 $B_d^0 \to f_1(1285) \pi^0$
&$1.4^{+0.4+0.6+0.5+0.0+0.2}_{-0.3-0.4-0.3-0.0-0.2}
\times 10^{-7}$
& $B_s^0 \to f_1(1285) \pi^0$
&$2.7^{+0.9+0.2+1.6+0.7+0.3}_{-0.7-0.2-1.3-0.5-0.2}
\times 10^{-8}$
 \\
 $B_d^0 \to f_1(1420) \pi^0$
&$1.1^{+0.0+0.6+0.4+0.3+0.1}_{-0.1-0.3-0.1-0.2-0.1}
\times 10^{-8}$
& $B_s^0 \to f_1(1420) \pi^0$
&$1.4^{+0.5+0.1+0.8+0.1+0.1}_{-0.4-0.1-0.6-0.1-0.1}
\times 10^{-7}$
 \\
 $B_d^0 \to f_1(1285) K^0$
&$1.8^{+0.5+1.3+2.1+0.3+0.2}_{-0.4-0.8-1.4-0.3-0.2}
\times 10^{-6}$
& $B_s^0 \to f_1(1285) \bar{K}^0$
&$7.4^{+2.7+0.6+6.6+2.0+2.5}_{-1.8-0.6-4.5-1.5-1.1}
\times 10^{-8}$
 \\
 $B_d^0 \to f_1(1420) K^0$
&$4.8^{+1.0+0.9+1.4+0.3+0.7}_{-0.8-0.7-1.2-0.3-0.5}
\times 10^{-6}$
& $B_s^0 \to f_1(1420) \bar{K}^0$
&$5.9^{+2.0+0.5+3.9+0.1+1.1}_{-1.4-0.5-2.9-0.2-0.8}
\times 10^{-7}$
 \\
 $B_d^0 \to f_1(1285) \eta$
&$1.0^{+0.2+0.5+0.2+0.0+0.3}_{-0.1-0.4-0.1-0.0-0.1}
\times 10^{-7}$
& $B_s^0 \to f_1(1285) \eta$
&$3.9^{+1.6+0.4+1.3+0.8+1.5}_{-1.0-0.4-1.2-0.7-0.9}
\times 10^{-6}$
 \\
 $B_d^0 \to f_1(1420) \eta$
&$1.7^{+0.2+0.9+0.9+0.5+0.1}_{-0.2-0.6-0.7-0.4-0.0}
\times 10^{-8}$
& $B_s^0 \to f_1(1420) \eta$
&$1.3^{+0.4+0.1+0.5+0.1+0.3}_{-0.3-0.1-0.4-0.1-0.2}
\times 10^{-5}$
 \\
 $B_d^0 \to f_1(1285) \eta'$
&$3.3^{+0.2+1.8+1.6+0.2+0.9}_{-0.1-1.2-1.1-0.2-0.2}
\times 10^{-8}$
& $B_s^0 \to f_1(1285) \eta'$
&$3.4^{+1.3+0.5+0.4+0.6+0.9}_{-0.9-0.4-0.4-0.5-0.6}
\times 10^{-6}$
 \\
 $B_d^0 \to f_1(1420) \eta'$
&$5.0^{+0.9+0.8+2.6+0.2+1.0}_{-0.6-0.6-2.0-0.2-0.8}
\times 10^{-8}$
& $B_s^0 \to f_1(1420) \eta'$
&$1.1^{+0.2+0.1+0.1+0.1+0.3}_{-0.2-0.1-0.1-0.1-0.2}
\times 10^{-5}$
 \\
 \hline \hline
\end{tabular}}
\end{center}
\end{table}

\item[(3)]
By careful analysis on the decay amplitudes, it is found that the $B^+ \to f_1 \pi^+(\Delta S =0)$ decays are almost dominated by the contributions from factorizable emission diagrams.
Moreover, based on Eqs.~(\ref{eq:mix-f1-f1p}), (\ref{eq:f1285pip}), and
(\ref{eq:f1285pi-d}), and the numerical results
of the branching ratios in Table~\ref{tab:BR-Bu}, one can straightforwardly see the constructive (destructive)
effects to the $B_{u,d} \to f_1(1285) \pi\ [f_1(1420) \pi]$ decays.

Theoretically, these four decays have also been studied in the
QCDF,~\footnote{As stressed in the Introduction, the branching ratios of the $B \to f_1 P$ decays  given in the naive factorization are very crude. Thus we will only compare our predictions
with that obtained in the QCDF theoretically.}
and the numerical results can be read as(in units of $10^{-6}$)~\cite{Cheng:2007mx}
   \beq
Br(B^+ \to f_1(1285) \pi^+) &=&
\left\{ \begin{array}{ll}
5.2^{+1.5}_{-1.0}& \vspace{0.1cm} \\
4.6^{+1.3}_{-0.9}& \\ \end{array} \right. ,\quad\;\;\;
Br(B^0 \to f_1(1285) \pi^0) =
\left\{ \begin{array}{ll}
0.26^{+0.32}_{-0.11}& \vspace{0.1cm} \\
0.20^{+0.27}_{-0.09}& \\ \end{array} \right. \;;
\label{eq:f1285pi-QCDF}\\
Br(B^+ \to f_1(1420) \pi^+) &=&
\left\{ \begin{array}{ll}
0.06^{+0.01}_{-0.00}& \vspace{0.1cm} \\
0.59^{+0.21}_{-0.15}&  \\ \end{array} \right. ,\quad
Br(B^0 \to f_1(1420) \pi^0) =
\left\{ \begin{array}{ll}
0.003^{+0.005}_{-0.003}& \vspace{0.1cm} \\
0.05^{+0.05}_{-0.03}&  \\ \end{array} \right.\;.
\label{eq:f1420pi-QCDF}
   \eeq
Note that the predictions of the branching ratios for $B_{u,d} \to f_1 \pi$ decays
in the QCDF correspond to two different sets of $\theta_{^3\!P_1}$ in the flavor
singlet-octet basis, i.e., $27.9^\circ$(first entry) and $53.2^\circ$(second entry).
One can easily find the good agreement between the pQCD predictions with $\phi_{f_1} \sim 24^\circ$ and the QCDF predictions with $\theta_{^3\!P_1} \sim 53.2^\circ$
for the $B_{u,d} \to f_1 \pi$ decays within errors.

According to Ref.~\cite{Cheng:2013cwa},
the mixing of the $f_1(1285)-f_1(1420)$ system in the singlet-octet and quark-flavor
bases can be written as the following form,
\begin{eqnarray}
 \left( \begin{array}{c}
    |f_1(1285) \rangle \\
    |f_1(1420)   \rangle \end{array} \right ) =
\left( \begin{array}{cc}
     \cos\theta_{^3\!P_1} & \sin\theta_{^3\!P_1}   \\
   -\sin\theta_{^3\!P_1}  & \cos\theta_{^3\!P_1} \end{array} \right )
    \left( \begin{array}{c}
                 |f_1 \rangle \\
                 |f_8 \rangle \end{array} \right )=
\left( \begin{array}{cc}
     \cos\alpha_{^3\!P_1} & \sin\alpha_{^3\!P_1}   \\
   -\sin\alpha_{^3\!P_1}  & \cos\alpha_{^3\!P_1} \end{array} \right )
    \left( \begin{array}{c}
                 |f_{1q} \rangle \\
                 |f_{1s} \rangle \end{array} \right )  \ ,
\end{eqnarray}
where $f_1$ and $f_8$ are the flavor singlet and flavor octet, respectively,
and the mixing angle $\alpha_{^3\!P_1}$ in the quark-flavor basis satisfies the
relation $\alpha_{^3\!P_1}= 35.3^\circ - \theta_{^3\!P_1}$ and measures the
deviation from ideal mixing. Then the $\alpha_{^3\!P_1} \sim -17.9^\circ$ can be
derived from the second entry in the QCDF, which thus leads to the same mixing form
as that adopted in this work, i.e., Eq.~(\ref{eq:mix-f1-f1p}) with a positive
value of the mixing angle.

Furthermore, a reasonable deduction obtained more naturally
is that the $f_1(1285)\ [f_1(1420)]$ is basically determined by the component $f_{1q}\ [f_{1s}]$
based on the following ratios(central values) between the branching
ratios of $B^+ \to f_1 \pi^+$ decays in the pQCD and QCDF approaches,
 \beq
R_{f_1(1285)\pi} &\equiv&
\frac{Br(B^+ \to f_1(1285) \pi^+)_{\rm QCDF}}{Br(B^+ \to f_1(1285) \pi^+)_{\rm pQCD}}
\approx  1.15 \sim |\frac{\cos{\alpha_{^3\!P_1}}}{\cos{\phi_{f_1}}}|^2 \approx 1.09\;,\\
R_{f_1(1420)\pi} &\equiv&
\frac{Br(B^+ \to f_1(1420) \pi^+)_{\rm QCDF}}{Br(B^+ \to f_1(1420) \pi^+)_{\rm pQCD}}
\approx  0.80 \sim |\frac{\sin{\alpha_{^3\!P_1}}}{\sin{\phi_{f_1}}}|^2 \approx 0.58\;.
 \eeq
Notice that the above relations cannot be easily deduced from the $B_d^0 \to f_1 \pi^0$
modes. The underlying reason is that the former $B^+ \to f_1 \pi^+$ decays are with
the dominant tree(color-allowed) contributions and negligible penguin pollution, while
the latter $B_d^0 \to f_1
\pi^0$ channels embrace the small tree(color-suppressed) and more important penguin
contributions.  The predictions of the \cp-averaged branching ratios for
$B_{u,d} \to f_1 \pi$ decays in the pQCD approach with the
corresponding phenomenological discussions
are expected to be tested by the near future experiments at LHC.

\item[(4)]
For the penguin-dominated $B_{u,d} \to f_1 K$ decays, the destructive (constructive)
interferences between $f_{1q} K$ and $f_{1s} K$ result in the approximately
equal branching ratios for $B_{u,d} \to f_1(1285) K\ [f_1(1420) K]$ decays,
\beq
Br(B^+ \to f_1(1285) K^+)&=& 1.6^{+2.2}_{-1.4}\times 10^{-6}  \sim
Br(B_d^0 \to f_1(1285) K^0) = 1.8^{+2.5}_{-1.7} \times 10^{-6} \;, \\
Br(B^+ \to f_1(1420) K^+)&=&  5.1^{+2.1}_{-1.7}\times 10^{-6}   \sim
Br(B_d^0 \to f_1(1420) K^0) = 4.8^{+2.1}_{-1.7} \times 10^{-6} \;,
\eeq
which indicate that the tree contributions are highly suppressed because of
$|\lambda_u^s|:|\lambda_t^s| \sim 0.02$.
Of course, it is worth stressing that, in terms of the central values of the decay rates, the color-allowed tree
contributions(around $10\%$) of $B^+ \to f_1 K^+$ decays are larger than those color-suppressed ones (almost $0\%$) of $B_d^0 \to f_1 K^0$ decays, though which are negligible relative to dominant penguin contributions in both sets of decay modes.

The predictions on the branching ratios have also been presented in the framework of QCDF(in
units of $10^{-6}$)~\cite{Cheng:2007mx}:
\beq
Br(B^+ \to f_1(1285) K^+) &=& 5.2^{+9.7}_{-10.1} \;, \quad
Br(B^+ \to f_1(1420) K^+) = 13.8^{+18.4}_{-7.8} \;,\\
Br(B_d^0 \to f_1(1285) K^0) &=&  5.2^{+4.5}_{-2.2}\;, \quad\;\;\;
Br(B_d^0 \to f_1(1420) K^0) = 13.1^{+17.5}_{-7.3} \;.
\eeq
In view of the better consistency observed from the $B_{u,d} \to f_1 \pi$ decays
theoretically, we here only quote the second entry of the branching ratios
for $B_{u,d} \to f_1 K$ decays in the QCDF for clarification.
It can be seen that
the theoretical predictions in both pQCD and QCDF approaches are basically
consistent with each other within still large uncertainties. However,
as far as the central values are considered, $Br(B_{u,d} \to f_1 K)_{\rm QCDF}$
are a bit larger than $Br(B_{u,d} \to f_1 K)_{\rm pQCD}$ with a factor
near 3.

As mentioned in the Introduction, there are just the preliminary upper limits
of branching ratios for the $B^+ \to f_1 K^+$ decays made by the {\it BABAR}
Collaboration~\cite{Burke:2007zz},
\beq
Br(B^+ \to f_1(1285) K^+) &<&   2.0 \times 10^{-6}\;,
\eeq
and
\beq
Br(B^+ \to f_1(1420) K^+)\cdot Br(f_1(1420) \to \bar K^* K) &<&
4.1 \times 10^{-6}\;,  \\
Br(B^+ \to f_1(1420) K^+)\cdot Br(f_1(1420) \to \eta \pi\pi) &<&
2.9 \times 10^{-6}\;.
\eeq
We can find that the prediction for $Br(B^+ \to f_1(1285) K^+)$
in the pQCD approach is in good agreement
with the preliminary upper limit, while that in the QCDF is barely
consistent with the experimental limit within large theoretical errors.
There are no 
accurate values of the decay rates of $f_1(1420) \to \bar K^* K$
and $\eta \pi \pi$ modes currently, which consequently results in no available upper bound
for $B^+ \to f_1(1420) K^+$ channel.
But, it can be imagined that we can extract phenomenologically the information
on the decay rates of
$f_1(1420) \to \bar K^* K$ and $\eta \pi\pi$ decays if our predictions
of $Br(B^+ \to f_1(1420) K^+) \sim {\cal O}(10^{-6})$ are 
confirmed by the measurements
at LHCb and Belle II 
experiments in the near future. Of course, we first need to await
enough data samples to test our theoretical predictions.

In order to observe the dependence on the mixing angle $\phi_{f_1}$ of the
$B^+ \to f_1 K^+$ decays, we simply examine the central values
of the branching ratios in the pQCD approach as a function of $\phi_{f_1}$
in the range of $[0, 90^\circ]$,
which can be seen in Fig.~\ref{fig:fig2}. One can observe that
the $\phi_{f_1}$ dependence of the $B^+ \to f_1(1420) K^+$ mode is opposite
to that of the $B^+ \to f_1(1285) K^+$ directly from Fig.~\ref{fig:fig2}.
\begin{figure}[!t]
\begin{center}
\hspace{-1 cm}
\includegraphics[scale=0.6]{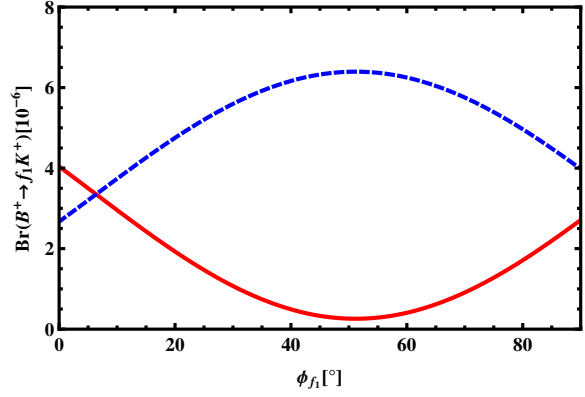}
\caption{(Color online) Dependence on the mixing angle $\phi_{f_1}$
 of the branching ratios of $B^+ \to f_1 K^+$ in the pQCD approach.
The red solid\ [blue dashed] line corresponds to the $B^+ \to f_1(1285) K^+\ [f_1(1420) K^+]$
decay, respectively.  }
\label{fig:fig2}
\end{center}
\end{figure}
Moreover, we also present the branching ratios of $B^+ \to f_1 K^+$ decays in the pQCD
approach with $\phi_{f_1} \sim 15^\circ$ and $20^\circ$ as a reference:
\beq
Br(B^+ \to f_1(1285) K^+) &=& 2.4^{+2.9}_{-1.9}\times 10^{-6} \;, \quad
Br(B^+ \to f_1(1420) K^+) = 4.3^{+1.6}_{-1.4}\times 10^{-6} \;,
\eeq
with $\phi_{f_1} \sim 15^\circ$ and
\beq
Br(B^+ \to f_1(1285) K^+) &=& 1.9^{+2.6}_{-1.6}\times 10^{-6} \;, \quad
Br(B^+ \to f_1(1420) K^+) = 4.8^{+1.7}_{-1.7}\times 10^{-6} \;,
\eeq
with $\phi_{f_1} \sim 20^\circ$.
According to the brief review of the $f_1(1285)-f_1(1420)$ mixing in the last section,
in terms of the central values of the currently existing mixing angle $\phi_{f_1}$
from both theoretical and experimental sides, one can find that the angle $\phi_{f_1}$
lies in the range of $[15^\circ, 27^\circ]$~\cite{Cheng:2013cwa}. Similarly, if the
preliminary upper limit
for the branching ratio of the $B^+ \to f_1(1285) K^+$ mode could be
considered as the central value of the experimental measurement, then we can find a rough constraint
of the mixing angle $\phi_{f_1}$ through the numerical evaluations
in the pQCD approach, i.e., $\phi_{f_1} \in [20^\circ, 27^\circ]$.

\item[(5)]
The \cp-averaged branching ratios of $B_{d}^0 \to f_1 (\eta, \eta')$ decays
in the pQCD approach are presented in Table~\ref{tab:BR-Bds}.
As a matter of fact, it is noted that these decays include two sets of destructive and/or
constructive effects simultaneously
due to $\eta-\eta'$ mixing and $f_1(1285)-f_1(1420)$ mixing.
We find that $Br(B_d^0 \to f_1(1285) \eta) \sim 5 \times
Br(B_d^0 \to f_1(1420) \eta)$ and $Br(B_d^0 \to f_1(1285) \eta')
\sim Br(B_d^0 \to f_1(1420) \eta')$ within errors. While in terms of their
central values of the branching ratios, 
we can easily find the constructive (destructive) interferences in $B_d^0 \to f_1(1285) \eta\ [f_1(1420) \eta]$ decays and the slightly destructive (constructive) effects in
$B_d^0 \to f_1(1285) \eta'\ [f_1(1420) \eta']$ ones.
And the similarly interesting phenomena
can be found correspondingly
in $B_d^0 \to f_1(1285) \eta [\eta']$ and $B_d^0 \to f_1(1420) \eta [\eta']$ decays.
Because of the similar behavior in both vector and axial-vector
mesons and this interesting pattern also occurring in the $B_d^0 \to (\omega, \phi) (\eta, \eta')$ decays~\cite{Wang:2005bk}, it is reasonable to conjecture that the $f_1(1285)\ [f_1(1420)]$ is
dominated by the $f_{1q}\ [f_{1s}]$.
However, all magnitudes of these four branching ratios are so small that the
current experiments cannot observe them in a short period, which then have to be detected in the future.

\item[(6)]
To our best knowledge, the $B_s^0 \to f_1 P$ decays are studied for the first
time in the pQCD approach and their estimations on the physical observables such
as \cp-averaged branching ratios and \cp-violating asymmetries have been collected
in the Tables~\ref{tab:BR-Bds} and \ref{tab:d-m-cp-bds}.

\begin{itemize}

\item[(a)]
As shown in Table~\ref{tab:BR-Bds},  the \cp-averaged branching ratios of $B_s^0 \to f_1 (\pi^0, \bar K^0)$ decays are very small, around the order of $10^{-8}
\sim 10^{-7}$ in the pQCD approach, which cannot be easily reached in the near future
experiments.
Relative to $B_d^0 \to f_1 K^0$ decays, the $B_s^0 \to f_1 \bar K^0$ ones
are also the penguin-dominated processes with dramatically small tree amplitudes through the
$\Delta S =0$ transitions. Due to the CKM hierarchy, the moduli of $\lambda_t^d$ is just about 22\% of that of $\lambda_t^s$, which consequently leads to $Br(B_s^0 \to f_1
\bar K^0) < Br(B_d^0 \to f_1 K^0)$ as naive expectations.
Different from $B_d^0 \to f_1 \pi^0$ decays, the $B_s^0 \to f_1 \pi^0$ decays have no
$B_s^0 \to \pi^0$ transitions
and are nearly determined by the factorizable emission contributions via $B_s^0 \to f_{1s}$
transitions. Based on Eq.~(\ref{eq:mix-f1-f1p}), the coefficients $-\sin\phi_{f_1}$ and
$\cos\phi_{f_1}$ can be found in the $B_s^0 \to f_1(1285) \pi^0$ and $B_s^0 \to f_1(1420)
\pi^0$ decays, respectively, which thus result in the smaller (larger) branching ratio
of the former (latter) mode with $\sin^2(24^\circ) \sim 0.17\ [\cos^2(24^\circ)\sim 0.83]$.
The similar (contrary) decay pattern between $B_s^0 \to f_1 \bar K^0$ and
$B_d^0 \to f_1 K^0$[$B_s^0 \to f_1 \pi^0$
and $B_d^0 \to f_1 \pi^0$] modes can also clearly be seen from
Table~\ref{tab:BR-Bds}.

\item[(b)]

The \cp-averaged branching ratios of $B_s^0 \to f_1 (\eta, \eta')$ modes
completely dominated by the penguin contributions in the pQCD approach
are large, in the order of
$10^{-6} \sim 10^{-5}$, and are expected to be easily accessed by
the ongoing LHCb and forthcoming
Belle II 
experiments.
Without the so-called tree contaminations,
the central values of the decay rates of these four channels remain
unchanged in the pQCD approach as presented in Table~\ref{tab:BR-Bds}.
Similar to $B_d^0 \to f_1 (\eta, \eta')$
decays, the $B_s^0 \to f_1 (\eta, \eta')$ ones also embrace two sets of constructive
and/or destructive interferences because of the $\eta_q-\eta_s$ mixing and
$f_{1q}-f_{1s}$ mixing.
But, in contrast to the decay pattern of $B_d^0 \to f_1 (\eta, \eta')$,
as far as the central values are considered, we find the weakly constructive 
(destructive) effects to the $B_s^0 \to f_1(1285) \eta [\eta']$ and $B_s^0 \to f_1(1420)
\eta [\eta']$ decays, and the strongly destructive (constructive) interferences
in the $B_s^0 \to f_1(1285) \eta\ [f_1(1420) \eta]$ and $B_s^0 \to f_1(1285) \eta'\ [f_1(1420) \eta']$
ones. By considering the theoretical errors, we can obtain the relations $Br(B_s^0 \to
f_1(1285) \eta) \sim Br(B_s^0 \to
f_1(1285) \eta') \sim {\cal O}(10^{-6})$ and $Br(B_s^0 \to
f_1(1420) \eta) \sim Br(B_s^0 \to
f_1(1420) \eta') \sim {\cal O}(10^{-5})$ approximately.
It is therefore of great interest to examine these $B_s^0 \to f_1 (\eta, \eta')$ decays, 
with $10^{-6}$ and even larger branching ratios, 
and interesting phenomenologies at the
experimental aspects.

\end{itemize}

\item[(7)]
We also explore some ratios of the \cp-averaged branching ratios of the considered
$B \to f_1 P$ decays in the pQCD approach. For simplicity, we just present the ratios of
decay modes with large branching ratios. Therefore, the relevant ratios
can be read as follows:
\beq
R_1 &\equiv&  \frac{Br(B^+ \to f_1(1420) \pi^+)}
{Br(B^+ \to f_1(1285) \pi^+)} = 0.18^{+0.20}_{-0.16}\;, \\
R_2 &\equiv&  \frac{Br(B^+ \to f_1(1285) K^+)}
{Br(B^+ \to f_1(1420) K^+)} = 0.31^{+0.45}_{-0.29}\;, \\
R_3 &\equiv&  \frac{Br(B_d^0 \to f_1(1285) K^0)}
{Br(B_d^0 \to f_1(1420) K^0)} = 0.38^{+0.55}_{-0.38}\;, \\
R_4 &\equiv&  \frac{Br(B_s^0 \to f_1(1285) \eta)}
{Br(B_s^0 \to f_1(1420) \eta)} = 0.30^{+0.26}_{-0.21}\;, \\
R_5 &\equiv&  \frac{Br(B_s^0 \to f_1(1285) \eta')}
{Br(B_s^0 \to f_1(1420) \eta')} = 0.31^{+0.20}_{-0.15}\;.
\eeq
One can directly observe that the ratio $R_1$ from $B^+ \to f_1 \pi^+(\Delta S=0)$
is very different from the other four similar ratios $R_{2-5}$ from $B_{u,d} \to f_1 K
(\Delta S =1)$ and $B_s^0 \to f_1 (\eta, \eta')(\Delta S =1)$.
The measurements of these ratios
will be helpful to understand the mixing angle $\phi_{f_1}$
of the $f_1(1285)-f_1(1420)$ system effectively and further
determine the definite components of both $f_1$ mesons.

\item[(8)]
As mentioned in the Introduction, the contributions from
weak annihilation diagrams play important roles in the heavy $B$ meson decays, which are
complex with a sizable strong phase proposed by the pQCD approach and supported by the
QCDF approach through fitting to the data, although the contrary viewpoint has been
stated by soft-collinear effective theory. We will therefore analyze the annihilation
contributions in these 20 $B \to f_1 P$ decays. For the sake of simplicity,
we here will only take the central values of the branching ratios in the pQCD approach for clarification.

\begin{itemize}
\item[(a)]
For the $\Delta S =0 $ processes, when the weak annihilation contributions are neglected,
then the branching ratios of the ten $b \to d$ transitions can be presented as follows:
\beq
Br(B^+ \to f_1(1285) \pi^+) &= &  3.9 \times 10^{-6}\;,  \quad\;\;\;
Br(B^+ \to f_1(1420) \pi^+) =   7.1 \times 10^{-7}\;; \\
Br(B_d^0 \to f_1(1285) \pi^0) &= & 1.2 \times 10^{-7} \;,  \quad\;\;\;\;\
Br(B_d^0 \to f_1(1420) \pi^0) =   2.1 \times 10^{-9}\;; \\
Br(B_d^0 \to f_1(1285) \eta) &= &  8.7 \times 10^{-8}\;,  \quad\;\;\;\;\;\;\;
Br(B_d^0 \to f_1(1420) \eta) =   5.1 \times 10^{-9}\;; \\
Br(B_d^0 \to f_1(1285) \eta') &= &  4.3 \times 10^{-9}\;,  \quad\;\;\;\;\;\;
Br(B_d^0 \to f_1(1420) \eta') =   4.3 \times 10^{-9} \;; \\
Br(B_s^0 \to f_1(1285) \bar K^0) &= &  5.5 \times 10^{-8}\;,  \quad\;\;\;\;
Br(B_s^0 \to f_1(1420) \bar K^0) =   5.3 \times 10^{-7}\;.
\eeq

\item[(b)]
For the $\Delta S =1 $ channels, when the weak annihilation contributions are turned off,
then the decay rates of the ten $b \to s$ transitions can be given as follows:
\beq
Br(B^+ \to f_1(1285) K^+) &= &  1.8 \times 10^{-6}\;,  \quad\;\;\;
Br(B^+ \to f_1(1420) K^+) =   3.5 \times 10^{-6}\;; \\
Br(B_d^0 \to f_1(1285) K^0) &= & 2.0 \times 10^{-6} \;,  \quad\;\;\;\;\
Br(B_d^0 \to f_1(1420) K^0) =   3.5 \times 10^{-6}\;; \\
Br(B_s^0 \to f_1(1285) \eta) &= &  4.2 \times 10^{-6}\;,  \quad\;\;\;\;\;\;\;\;
Br(B_s^0 \to f_1(1420) \eta) =   1.2 \times 10^{-5}\;; \\
Br(B_s^0 \to f_1(1285) \eta') &= &  3.4 \times 10^{-6}\;,  \quad\;\;\;\;\;\;\;
Br(B_s^0 \to f_1(1420) \eta') =   7.8 \times 10^{-6} \;; \\
Br(B_s^0 \to f_1(1285) \pi^0) &= &  2.7 \times 10^{-8}\;,  \quad\;\;\;\;\;\;
Br(B_s^0 \to f_1(1420) \pi^0) =   1.3 \times 10^{-7}\;.
\eeq

\end{itemize}
Compared with the values listed in Tables~\ref{tab:BR-Bu} and \ref{tab:BR-Bds},
one can find that the decays such as $B^+ \to f_1 \pi^+, f_1(1285) K^+$, $B_d^0 \to f_1(1285)
(\pi^0, K^0, \eta)$, and $B_s^0 \to f_1(1285) (\pi^0, \eta, \eta'),
f_1(1420) (\pi^0, \bar K^0, \eta')$ are not significantly sensitive to the weak
annihilation contributions. However, it is important to note that the modes such as
$B_{u,d} \to f_1(1420) K$, $B_d^0 \to f_1(1285) \eta', f_1(1420) (\pi^0, \eta, \eta')$, and
$B_s^0 \to f_1(1285) \bar K^0, f_1(1420) \eta'$ suffer from sizable annihilation effects;
specifically, without the contributions from annihilation diagrams, the branching
ratios decrease correspondingly by around 30\% for $B_{u,d} \to f_1(1420) K$ and $B_s^0
\to f_1(1420) \eta'$, $70\% \sim 90\%$ for $B_d^0 \to f_1 \eta',
f_1(1420) (\pi^0, \eta)$, and 26\% for $B_s^0 \to f_1(1285) \bar K^0$, respectively.
Of course, the reliability
of the contributions from the annihilation diagrams to these considered decays calculated in the
pQCD approach will be carefully examined by the relevant experiments in the future.

\item[(9)]
Frankly speaking, as the most important inputs in the calculations of pQCD approach,
the currently less constrained light-cone distribution amplitudes of the axial-vector
$f_1$ mesons result in the theoretical predictions of the branching ratios
for the considered 20 $B \to f_1 P$ decays with relatively large
uncertainties, which are expected to be greatly improved by the
LQCD calculations and/or large numbers of
related experiments in the future.
For example, analogous to $\eta/\eta' \to \gamma \gamma$~\cite{Agashe:2014kda,Donoghue:1986wv},
one can fix the mixing angle $\phi_{f_1}$ and/or the decay constants of axial-vector $f_1$ mesons
through the measurements on the decay widths of $f_1(1285)/f_1(1420) \to \gamma \gamma^*$
channels~\cite{Gidal:1987bn}. Of course, one can also determine the mixing angle $\phi_{f_1}$
through the Gell-Mann$-$Okubo mass formula for the $^3\!P_1$
axial-vector states, the relation of the decay rates of the radiative
$f_1(1285) \to \rho \gamma$ and $\phi \gamma$ modes or of the radiative $J/\psi \to f_1(1285) \gamma$
and $f_1(1420) \gamma$ processes, and so on.

\end{enumerate}

\subsection{\cp-violating asymmetries of $B \to f_1 P$ decays in the pQCD approach}

\begin{table}[hbt]
\caption{ The direct \cp\  violations $\acp^{\rm dir}$ for $B^+ \to
f_1 (\pi^+, K^+)$ decays in the pQCD approach. Apart from the last error
induced by the variations of CKM parameters $\bar \rho$ and $\bar \eta$, the
sources of the main uncertainties have been specified in the discussions
of \cp-averaged branching ratios. }
\label{tab:dcp-bu}
 \begin{center}\vspace{-0.3cm}{
\begin{tabular}[t]{c|c}
\hline  \hline
  Channels   &  direct \cp\ violations($\%$)  \\
   \hline
 $B^+ \to f_1(1285) \pi^+$
&$\hspace{0.22cm}
18.3^{+2.0}_{-1.9} (\omega_b)
^{+0.3}_{-0.4} (f_{f_1})
^{+3.3}_{-2.0}(a_i^M)
^{+0.2}_{-0.2} (\phi_{f_1})
^{+2.7}_{-2.0} (a_t)
^{+0.7}_{-1.3}(V_i)
$
 \\
 $B^+ \to f_1(1420) \pi^+$
&$\hspace{0.22cm}
28.2^{+2.8}_{-2.8} (\omega_b)
^{+1.8}_{-1.4} (f_{f_1})
^{+5.8}_{-4.0}(a_i^M)
^{+1.1}_{-1.0} (\phi_{f_1})
^{+2.7}_{-2.4} (a_t)
^{+1.1}_{-1.7}(V_i)
$
 \\
 $B^+ \to f_1(1285) K^+$
&$-21.2^{+1.6}_{-1.9} (\omega_b)
^{+4.3}_{-1.9} (f_{f_1})
^{+12.8}_{-24.0}(a_i^M)
^{+2.4}_{-1.3} (\phi_{f_1})
^{+0.1}_{-0.0} (a_t)
^{+0.8}_{-1.3}(V_i)
$
 \\
 $B^+ \to f_1(1420) K^+$
&$-13.6^{+0.6}_{-0.5} (\omega_b)
^{+1.5}_{-1.3} (f_{f_1})
^{+2.3}_{-2.0}(a_i^M)
^{+0.9}_{-0.9} (\phi_{f_1})
^{+0.5}_{-0.4} (a_t)
^{+0.6}_{-0.5}(V_i)
$
 \\
 \hline \hline
\end{tabular}}
\end{center}
\end{table}

Now we come to the evaluations of the \cp-violating asymmetries of $B
\to f_1 P$ decays in the pQCD approach.
For the charged $B^+ \to f_1 (\pi^+, K^+)$ decays, the direct \cp\ violation
$\acp^{\rm dir}$ can be defined as,
 \beq
\acp^{\rm dir} =  \frac{|\overline{\cal A}_f|^2 - |{\cal A}_f|^2}{
 |\overline{\cal A}_f|^2+|{\cal A}_f|^2},
\label{eq:acp1}
\eeq
where ${\cal A}_f$ stands for the decay amplitudes of $B^+ \to f_1 \pi^+$
and $B^+ \to f_1 K^+$, respectively, while $\ov{{\cal A}}_f$ denotes the
charge conjugation $B^- \to f_1 \pi^-$ and $B^- \to f_1 K^-$ ones correspondingly.
Using Eq.~(\ref{eq:acp1}), the pQCD predictions for the direct \cp-violating asymmetries
of $B^+ \to f_1 (\pi^+, K^+)$ modes have been collected in Table~\ref{tab:dcp-bu},
in which we can easily find the large direct \cp\ violations for the four charged
$B^+ \to f_1 \pi^+$ and $f_1 K^+$ decays within errors as follows:
\beq
\acp^{\rm dir}(B^+ \to f_1(1285) \pi^+) &=&\hspace{0.26cm} (18.3^{+4.8}_{-3.7}) \%  \;, \qquad\;\;
\acp^{\rm dir}(B^+ \to f_1(1420) \pi^+) =\hspace{0.35cm} (28.2^{+7.4}_{-6.0}) \% \;; \label{eq:dcp-f1pip} \\
\acp^{\rm dir}(B^+ \to f_1(1285) K^+) &=& (-21.2^{+13.8}_{-24.2}) \%  \;, \qquad
\acp^{\rm dir}(B^+ \to f_1(1420) K^+) =  (-13.6^{+3.1}_{-2.7}) \% \;, \label{eq:dcp-f1kp}
\eeq
where various errors from the variations of the input parameters have been added in quadrature.
These large direct \cp-violating asymmetries combined with the large \cp-averaged
branching ratios[${\cal O}(10^{-6})$] are believed to be clearly measurable at
the LHCb and Belle II 
experiments.

\begin{table}[b]
\caption{ The direct \cp\ asymmetries $\acp^{\rm dir}$(first entry) and the mixing-induced \cp\
asymmetries $\acp^{\rm mix}$(second entry) for $B_{d(s)}^0 \to f_1 (\pi^0, K^0, \eta, \eta')$
decays in the pQCD approach. Moreover, the third entry in the right-hand
side is for the observable $A_{\Delta\Gamma_s}$ in $B_s^0$ meson decays.
Various errors arising from the input parameters as specified in previous section
have been added in quadrature.}
\label{tab:d-m-cp-bds}
 \begin{center}\vspace{-0.3cm}{
\begin{tabular}[t]{c|l|c|l}
\hline  \hline
  Channels   &  \cp\ asymmetries($\%$)
& Channels   &  \cp\ asymmetries($\%$) \\
   \hline
 $B_d^0 \to f_1(1285) \pi^0$
&$\begin{array}{l}
\hspace{0.22cm}
70.8^{+12.5}_{-17.0}
 \\
\hspace{0.4cm}
2.6^{+38.9}_{-32.1}
\end{array} $
&  $B_s^0 \to f_1(1285) \pi^0$
&$\begin{array}{l}
-29.5^{+10.7}_{-13.3}
\\
\hspace{0.16cm}
-9.9^{+14.0}_{-12.0}
\\
\hspace{0.26cm}
95.0^{+2.8}_{-4.4}
\end{array} $
 \\
 \hline
 $B_d^0 \to f_1(1420) \pi^0$
&$\begin{array}{l}
\hspace{0.25cm}
42.0^{+51.8}_{-78.6}
\\
\hspace{0.3cm}
86.1^{+13.6}_{-48.8}
\end{array} $
& $B_s^0 \to f_1(1420) \pi^0$
& $\begin{array}{l}
\hspace{0.25cm}
14.3^{+10.3}_{-6.9}
\\
-11.3^{+12.3}_{-10.2}
\\
\hspace{0.26cm}
98.3^{+1.0}_{-1.8}
\end{array}
$\\
 \hline \hline
 $B_d^0 \to f_1(1285) K^0$
&$\begin{array}{l}
\hspace{0.4cm}
2.3^{+2.5}_{-1.3}
\\
\hspace{0.25cm}
70.0^{+3.1}_{-2.9}
\end{array} $
& $B_s^0 \to f_1(1285) \bar K^0$
&$\begin{array}{l}
\hspace{0.25cm}
26.0^{+34.9}_{-30.0}
\\
-70.9^{+36.3}_{-19.8}
\\
\hspace{0.26cm}
65.5^{+28.7}_{-41.0}
\end{array}
$
 \\
 \hline
 $B_d^0 \to f_1(1420) K^0$
&$\begin{array}{l}
\hspace{0.4cm}
0.6^{+0.4}_{-0.5}
\\
\hspace{0.25cm}
69.9^{+2.4}_{-2.2}
\end{array}$
& $B_s^0 \to f_1(1420) \bar K^0$
&$\begin{array}{l}
\hspace{0.16cm}
-2.9^{+3.7}_{-4.3}
\\
-67.9^{+6.1}_{-5.8}
\\
\hspace{0.26cm}
73.4^{+5.3}_{-6.0}
\end{array}
$
 \\
 \hline \hline
 $B_d^0 \to f_1(1285) \eta$
&$\begin{array}{l}
-80.9^{+29.0}_{-15.4}
\\
-42.2^{+48.2}_{-43.9}
\end{array}$
& $B_s^0 \to f_1(1285) \eta$
&$\begin{array}{l}
\hspace{0.40cm}
1.0^{+1.2}_{-0.9}
\\
\hspace{0.4cm}
0.9^{+1.4}_{-1.6}
\\
\hspace{0.35cm}
\sim 100
\end{array}
$
 \\
 \hline
 $B_d^0 \to f_1(1420) \eta$
&$\begin{array}{l}
-93.6^{+20.8}_{-9.2}
\\
-13.3^{+51.8}_{-48.0}
\end{array} $
& $B_s^0 \to f_1(1420) \eta$
&$\begin{array}{l}
\hspace{0.16cm}
-1.4^{+0.4}_{-0.5}
\\
\hspace{0.4cm}
0.3^{+0.9}_{-1.2}
\\
\hspace{0.35cm}
\sim 100
\end{array}
$
 \\
 \hline \hline
 $B_d^0 \to f_1(1285) \eta'$
&$\begin{array}{l}
-47.7^{+31.5}_{-25.5}
\\
-86.3^{+20.1}_{-13.0}
\end{array}$
& $B_s^0 \to f_1(1285) \eta'$
&$\begin{array}{l}
\hspace{0.16cm}
-2.5^{+0.9}_{-0.9}
\\
\hspace{0.16cm}
-1.2^{+1.4}_{-1.3}
\\
\hspace{0.35cm}
\sim 100
\end{array}
$
 \\
 \hline
 $B_d^0 \to f_1(1420) \eta'$
&$\begin{array}{l}
\hspace{0.25cm}
29.0^{+24.1}_{-24.6}
\\
-44.2^{+17.6}_{-16.5}
\end{array} $
& $B_s^0 \to f_1(1420) \eta'$
&$\begin{array}{l}
\hspace{0.40cm}
1.5^{+0.7}_{-0.8}
\\
\hspace{0.4cm}
0.5^{+1.0}_{-1.0}
\\
\hspace{0.35cm}
\sim 100
\end{array}
$
 \\
 \hline \hline
\end{tabular}}
\end{center}
\end{table}

As for the \cp-violating asymmetries for the neutral $B_{d(s)}^0 \to f_1 P$ decays, the effects of
$B_{d(s)}^0-\bar{B}_{d(s)}^0$ mixing should be considered. The
\cp-violating asymmetries of $B_{d(s)}^0(\bar{B}_{d(s)}^0) \to
f_1 (\pi^0, K^0, \eta, \eta')$ decays are time dependent and
can be defined as
\beq
\acp &\equiv& \frac{\Gamma\left
(\bar{B}_{d(s)}^0(\Delta t) \to f_{CP}\right) -
\Gamma\left(B_{d(s)}^0(\Delta t) \to f_{CP}\right )}{ \Gamma\left
(\bar{B}_{d(s)}^0(\Delta t) \to f_{CP}\right ) + \Gamma\left
(B_{d(s)}^0(\Delta t) \to f_{CP}\right ) }\non
&=& \acp^{\rm dir} \cos(\Delta m_{d(s)}  \Delta t)
+ \acp^{\rm mix} \sin (\Delta m_{d(s)} \Delta
t), \label{eq:acp-def}
\eeq
where $\Delta m_{d(s)}$ is the mass
difference between the two $B_{d(s)}^0$ mass eigenstates, $\Delta
t =t_{CP}-t_{tag} $ is the time difference between the tagged
$B_{d(s)}^0$ [$\bar{B}_{d(s)}^0$] and the accompanying
$\bar{B}_{d(s)}^0$ [$B_{d(s)}^0$] with opposite $b$ flavor
decaying to the final \cp-eigenstate $f_{CP}$ at the time $t_{CP}$.
The direct and mixing-induced \cp-violating asymmetries $\acp^{\rm
dir} ({\cal C}_f)$
and $\acp^{\rm mix} ({\cal S}_f)$ can be written as
\beq
\acp^{\rm dir}= {\cal C}_f = \frac{ \left |
\lambda_{\rm CP}\right |^2-1 } {1+|\lambda_{\rm CP}|^2}, \qquad
\acp^{\rm mix}={\cal S}_f= \frac{ 2 {\rm Im}
(\lambda_{\rm CP})}{1+|\lambda_{\rm CP}|^2},
\label{eq:acp-csf}
\eeq
with the \cp-violating parameter $\lambda_{\rm CP}$
\beq
\lambda_{\rm CP} &\equiv& \eta_f \; \frac{V_{tb}^*V_{td(s)}}{V_{tb}V_{td(s)}^*}
\cdot \frac{ \langle f_{CP} |H_{eff}|\bar{B}_{d(s)}^0\rangle}
{\langle f_{CP} |H_{eff}|B_{d(s)}^0\rangle},
\label{eq:lambda}
\eeq
where $\eta_f$ is the \cp-eigenvalue of the final states.
Moreover, for $B_s^0$ meson decays, a non-zero ratio $(\Delta
\Gamma/\Gamma)_{B_s^0}$ is expected in the
SM~\cite{Beneke:1998sy,Fernandez:2006qx}.
For $B_s^0 \to f_1 (\pi^0, \bar K^0, \eta, \eta')$
decays, the third term $A_{\Delta \Gamma_s}$ related
to the presence of a non-negligible $\Delta \Gamma_s$
 to describe the \cp\ violation can be defined as follows~\cite{Fernandez:2006qx}:
\beq
A_{\Delta \Gamma_s} &=& \frac{ 2 {\rm Re} ( \lambda_{\rm CP})}{1+|\lambda_{\rm CP}|^2}.
\label{eq:acp-dgs}
\eeq
The three quantities describing the \cp\ violation in $B_s^0$ meson decays shown in Eqs.~(\ref{eq:acp-csf}) and
(\ref{eq:acp-dgs}) satisfy the following relation,
\beq
 |\acp^{\rm dir}|^2+ |\acp^{\rm mix}|^2+ |A_{\Delta \Gamma_s}|^2 &=&
 1 \;.\label{eq:summation-cp}
\eeq
Then, with the decay amplitudes for $B_{d(s)}^0 \to f_1 (\pi^0, K_S^0(\bar{K}_S^0), \eta, \eta')$ decays
as shown in the last section and the definitions in the above Eqs.~(\ref{eq:acp-csf})
-(\ref{eq:acp-dgs}), the direct and mixing-induced \cp-violating asymmetries
have been calculated in the pQCD
approach within large theoretical errors and displayed in Table~\ref{tab:d-m-cp-bds}.
Some remarks are in order:

\begin{enumerate}

\item[(1)]
As observed clearly from
Table~\ref{tab:d-m-cp-bds}, almost all of the $b \to d$ transition processes have the
large direct \cp\ violations with still large uncertainties, while most of the $b \to s$ transition
ones get the very small direct \cp\ asymmetries except for $B_s^0 \to f_1 \pi^0$ modes.

\item[(2)]
The relation of $\acp^{\rm dir}(B_d^0 \to f_1(1285) K_S^0) \sim 4
\times \acp^{\rm dir}(B_d^0 \to f_1(1420) K_S^0)$ can be found straightforwardly
from Table~\ref{tab:d-m-cp-bds}. The underlying reason is with different
contributions from tree diagrams because of the dominance of the $f_{1s}\ (f_{1q})$ component
in $f_1(1420)\ [f_1(1285)]$ in the current mixing form.
The same explanation can also be counted for the relation
$\acp^{\rm dir}(B^+ \to f_1 K^+) \gg \acp^{\rm dir}(B_d^0 \to f_1 K_S^0)$ in magnitudes.
Of course, as emphasized in the item on the discussions
of the branching ratios of $B^+ \to f_1 K^+$ and $B_d^0 \to f_1 K_S^0$ decays,
the latter modes are more like purely penguin-dominated channels.

\item[(3)]
It is interesting to note that those decays associated with very small direct \cp\ violations
but with very large \cp-averaged branching ratios are almost purely penguin-dominated modes,
whose tree pollution is so tiny that the numerical values of the decay rates
remain unchanged when just the penguin contributions are taken into account.
Actually, these mentioned decays, i.e., $B_d^0 \to f_1 K_S^0$ and $B_s^0 \to f_1 (\eta, \eta')$,
are induced by the $b \to s q \bar q$ mediated transitions with $q =u, d, s$
at the quark level. For the latter modes, in principle, we can utilize the mixing-induced
\cp\ asymmetries to study the $B_s^0-\bar B_s^0$ mixing phase $\phi_s$. Unfortunately, however,
these predictions in the pQCD approach suffer from significantly large theoretical
errors arising from the much less constrained hadronic parameters. Therefore, this issue
have to be left for future studies when the effective constraints are available
from the experiments and/or nonperturbative techniques such as LQCD calculations.
In the next subsection, we will analyze the $B_d^0-\bar B_d^0$ mixing phase $\phi_d$ explicitly
through the $B_d^0 \to f_1 K_S^0$ modes.

\item[(4)]

The third \cp-asymmetric observables $A_{\Delta_{\Gamma_s}}$ for the $B_s^0$ meson decays
are also listed in Table~\ref{tab:d-m-cp-bds}, in which we can find near 100\% for most of
the $B_s^0$ decay modes within large errors, apart from the $B_s^0 \to f_1(1420) \bar K^0$ channel
around 70\%. These interesting predictions in the pQCD approach and the resultant
phenomenologies are expected to be examined by the highly precise measurements
at the running LHCb and forthcoming Belle II 
experiments in the future.

\end{enumerate}

\subsection{Information on CKM weak phases $\alpha, \beta$, and $\gamma$ from $B \to f_1 P$ decays}

\begin{figure}[!!hbt]
\begin{center}
\hspace{-1 cm}
\includegraphics[scale=0.55]{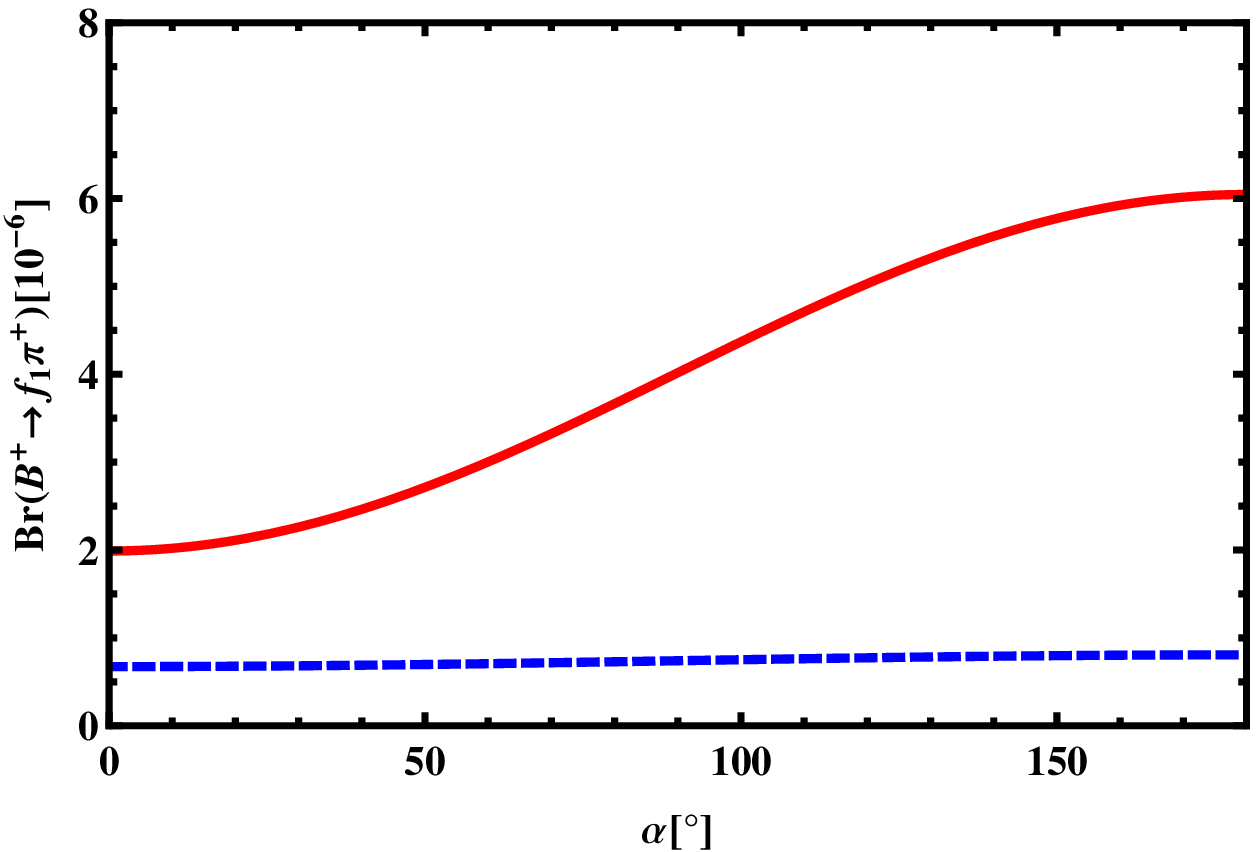}\hspace{0.6cm}
\includegraphics[scale=0.55]{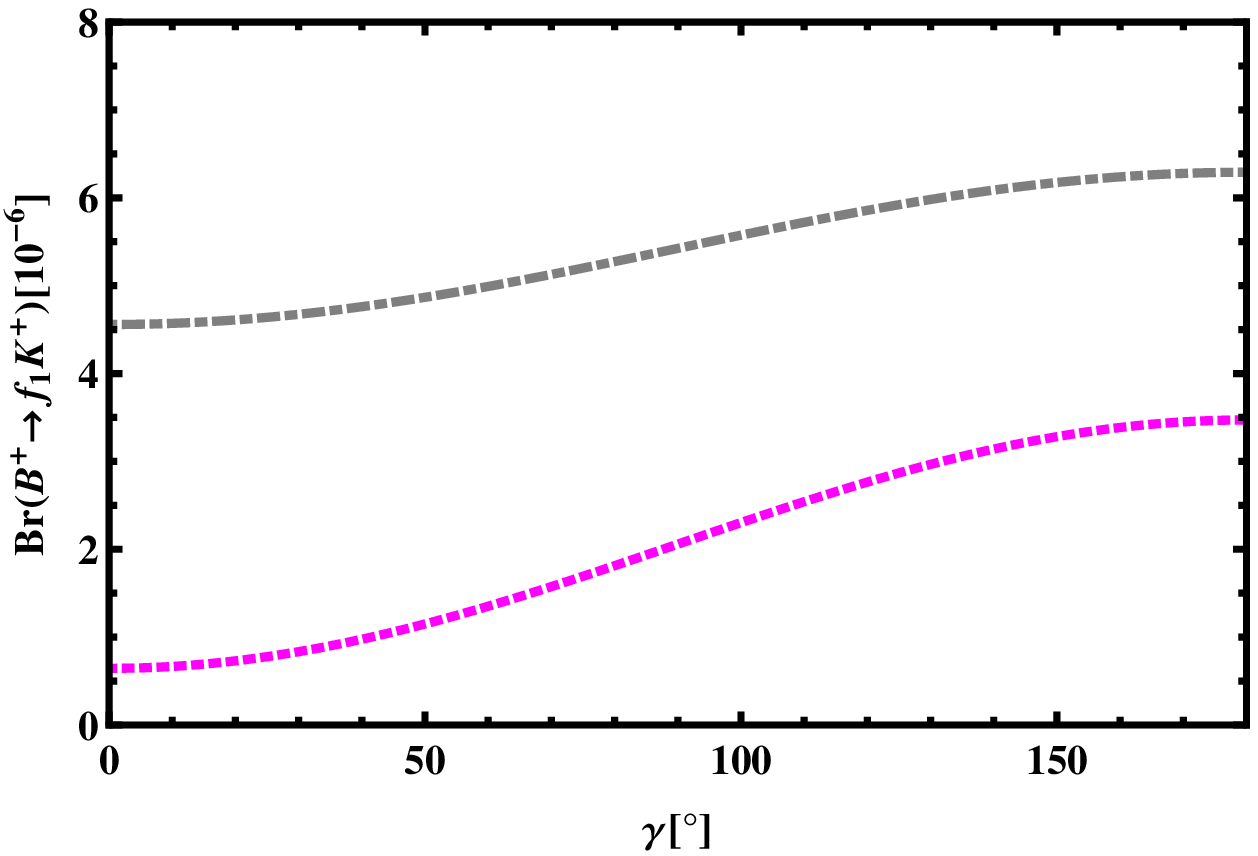}
\caption{(Color online) Dependence on the CKM weak phase $\alpha(\gamma)$ of central values
of the \cp-averaged branching ratios for $B^+ \to f_1 \pi^+(K^+)$ decays in the pQCD approach.
The red solid\ [blue dashed] line corresponds to the $B^+ \to f_1(1285) \pi^+\ [f_1(1420) \pi^+]$ decay
and the magenta dotted\ [gray dot-dashed] one corresponds to the $B^+ \to f_1(1285) K^+\ [f_1(1420) K^+]
$ mode, respectively. }
\label{fig:fig3}
\end{center}
\end{figure}

It is of great interest to note that the $B^+ \to f_1 \pi^+$ decays are $b \to d (\Delta S =0)$
transitions dominated by the tree diagrams, while the $B^+ \to f_1 K^+$ decays are
$b \to s (\Delta S =1)$ ones determined by the penguin contributions. These unique properties
exhibited in the $B^+ \to f_1 (\pi^+, K^+)$ decays motivate us to further explore more useful
information on the CKM weak phases $\alpha$ and $\gamma$ by employing
the careful investigations on the large \cp-averaged branching ratios and the large
direct \cp\ asymmetries of $B^+ \to f_1 (\pi^+, K^+)$ decays in the pQCD approach.

We know that the decay amplitudes
${\cal A}_f$ of $B^+ \to f_1 (\pi^+, K^+)$ can be further written as the following forms,
\beq
{\cal A}_f (B^+ \to f_1 \pi^+) &=& \lambda_u^d T - \lambda_t^d P =
\lambda_u^d T \{1 + r \exp{[i (\alpha + \delta)]}\} \;, \label{eq:M1A} \\
{\cal A}_f (B^+ \to f_1 K^+) &=& \lambda_u^s T' - \lambda_t^s P' =
\lambda_u^s T' \{1+ r' \exp{[i (\gamma' + \delta')]}\}\;, \label{eq:M1B}
\eeq
where $T(T')$ and $P(P')$ denote the tree and penguin decay amplitudes of
$B^+ \to f_1 \pi^+ (K^+)$ decays, and $r(r')$ and $\delta(\delta')$
represent the ratios of penguin to tree contributions $\frac{|\lambda_t^d||P|}{|\lambda_u^d||T|}
(\frac{|\lambda_t^s||P'|}{|\lambda_u^s||T'|})$  and the relative strong phases
between the corresponding tree and penguin diagrams.
The weak phase $\alpha$ come from the identity $\alpha = 180^\circ - \beta - \gamma$ with
the definitions $V_{td}=|V_{td}|\exp{(-i \beta)}$ and $V_{ub} = |V_{ub}|\exp{(-i \gamma)}$,
and the $\gamma'$ is defined as $\arg[{-\frac{V_{tb}^*V_{ts}}{V_{ub}^*V_{us}}}]$.
Then the decay amplitudes ${\cal \bar A}_f$ of the charge conjugated modes $B^- \to f_1
(\pi^-, K^-)$ can be easily written as
\beq
{\cal \bar A}_f (B^- \to f_1 \pi^-) &=& (\lambda_u^d)^* T - (\lambda_t^d)^* P =
(\lambda_u^d)^* T \{1 + r \exp{[i (-\alpha + \delta)]}\} \;, \label{eq:M1Ab}\\
{\cal \bar A}_f (B^- \to f_1 K^-) &=& (\lambda_u^s)^* T' - (\lambda_t^s)^* P' =
(\lambda_u^s)^* T' \{1+ r' \exp{[i (-\gamma' + \delta')]}\}\;.\label{eq:M1Bb}
\eeq
Therefore, the \cp-averaged branching ratios can be read as
\beq
Br(B^+ \to f_1 \pi^+) &\equiv& \frac{|{\cal \bar A}_f|_{f_1 \pi^-}^2 +|{\cal A}_f|_{f_1\pi^+}^2}{2} = |\lambda_u^d\; T|^2 \{ 1 + 2r\;\cos\alpha\cos\delta + r^2\} \;,
\label{eq:Br-1}\\
Br(B^+ \to f_1 K^+) &\equiv& \frac{|{\cal \bar A}_f|_{f_1 K^-}^2 +|{\cal A}_f|_{f_1K^+}^2}{2}
 = |\lambda_u^s\; T'|^2 \{ 1 + 2r'\cos\gamma\cos\delta' + r'^2\} \;,
 \label{eq:Br-2}
\eeq
in which $T^{(\prime)}, r^{(\prime)}$, and $\delta^{(\prime)}$
are all perturbatively calculated in the pQCD
approach; also, $\lambda_u^{d,s}$ are determined from the experiments. Thus, Eqs.~(\ref{eq:Br-1})
and (\ref{eq:Br-2}) can provide a possible way to determine the CKM angles $\alpha$ and $\gamma$ potentially by measuring the branching ratios, respectively. In Fig.~\ref{fig:fig3},
we show the central values of the
\cp-averaged branching ratios for $B^+ \to f_1(1285) \pi^+$(red solid line) and
 $B^+ \to f_1(1420) \pi^+$(blue dashed line) [$B^+ \to f_1(1285) K^+$(magenta dotted line) and
 $B^+ \to f_1(1420) K^+$(gray dot-dashed line)] decays as a function of the CKM weak phase $\alpha[\gamma]$ in the pQCD approach. One can easily see the strong (weak) dependence
on $\alpha$ for $B^+ \to f_1(1285) \pi^+\ [f_1(1420) \pi^+]$ decay and the moderate dependence
on $\gamma$ for $B^+ \to f_1 K^+$ decays in the pQCD approach from Fig.~\ref{fig:fig3}.
One can also directly observe from Fig.~\ref{fig:fig3} that the central
values of the branching ratios for the considered decays in the pQCD approach
correspond to the central values of $\alpha$ and $\gamma$ as around $89^\circ$ and
$70^\circ$, respectively, which are very consistent with the constraints from various
experiments~\cite{Agashe:2014kda}.

More information on the CKM angles $\alpha$ and $\gamma$ can also be hinted from
the large direct \cp\ asymmetries of $B^+ \to f_1 \pi^+(K^+)$ decays in the pQCD approach.
With Eqs.~(\ref{eq:M1A})$\sim$(\ref{eq:M1Bb}), the direct \cp-violating asymmetry
Eq.~(\ref{eq:acp1}) for $B^+ \to f_1 \pi^+ (K^+)$ can be described as the function of
$\alpha(\gamma)$,
\beq
\acp^{\rm dir}(B^+ \to f_1 \pi^+) &=& \frac{2 r\; \sin\alpha\; \sin\delta}{1
+2r\; \cos\alpha\; \cos\delta + r^2} \;,\label{eq:acp-r-d} \\
\acp^{\rm dir}(B^+ \to f_1 K^+) &=& -\frac{2 r'\; \sin\gamma\; \sin\delta'}{1
+2r'\; \cos\gamma\; \cos\delta' + {r'}^2} \;.\label{eq:acp-rp-dp}
\eeq
Again, as aforementioned, the ratios $r^{(\prime)}$ and the relative strong phases $\delta^{(\prime)}$
can be explicitly calculated in the pQCD approach. Undoubtedly, the former Eq.~(\ref{eq:acp-r-d}) is a function of $\sin\alpha$ and $\cos\alpha$,
and the latter Eq.~(\ref{eq:acp-rp-dp}) is a function of $\sin\gamma$ and $\cos\gamma$.
In particular, if one mode like $B^+ \to f_1(1420) \pi^+$ is almost completely tree dominated,
i.e., $r \ll 1$, then Eq.~(\ref{eq:acp-r-d}) can be further written approximately as
\beq
\acp^{\rm dir}(B^+ \to f_1(1420) \pi^+) &\sim&
2r\sin\alpha\sin\delta\;. \label{eq:acp-r-d-1}
\eeq
Analogously, if one mode like $B^+ \to f_1(1420) K^+$ is nearly pure penguin contributions,
i.e., $r' \gg 1$, then Eq.~(\ref{eq:acp-rp-dp}) can be further described approximately as
\beq
\acp^{\rm dir}(B^+ \to f_1 K^+) &\sim&
-\frac{2}{r'}\sin\gamma\sin\delta' \;.\label{eq:acp-rp-dp-1}
\eeq
Thus, the large direct \cp-violating asymmetries driven by these two equations, i.e.,
Eqs.~(\ref{eq:acp-r-d-1}) and (\ref{eq:acp-rp-dp-1}), will give rise
to the effective constraints more easily on the CKM phases $\alpha$ and $\gamma$ from the
experimental data with high precision. Certainly, based on Eqs.~(\ref{eq:acp-r-d-1})
and (\ref{eq:acp-rp-dp-1}), the large strong phases $\delta$ and $\delta'$ required
by the large direct \cp\ asymmetries can also be deduced naturally.

\begin{figure}[!!t]
\begin{center}
\hspace{-1 cm}
\includegraphics[scale=0.55]{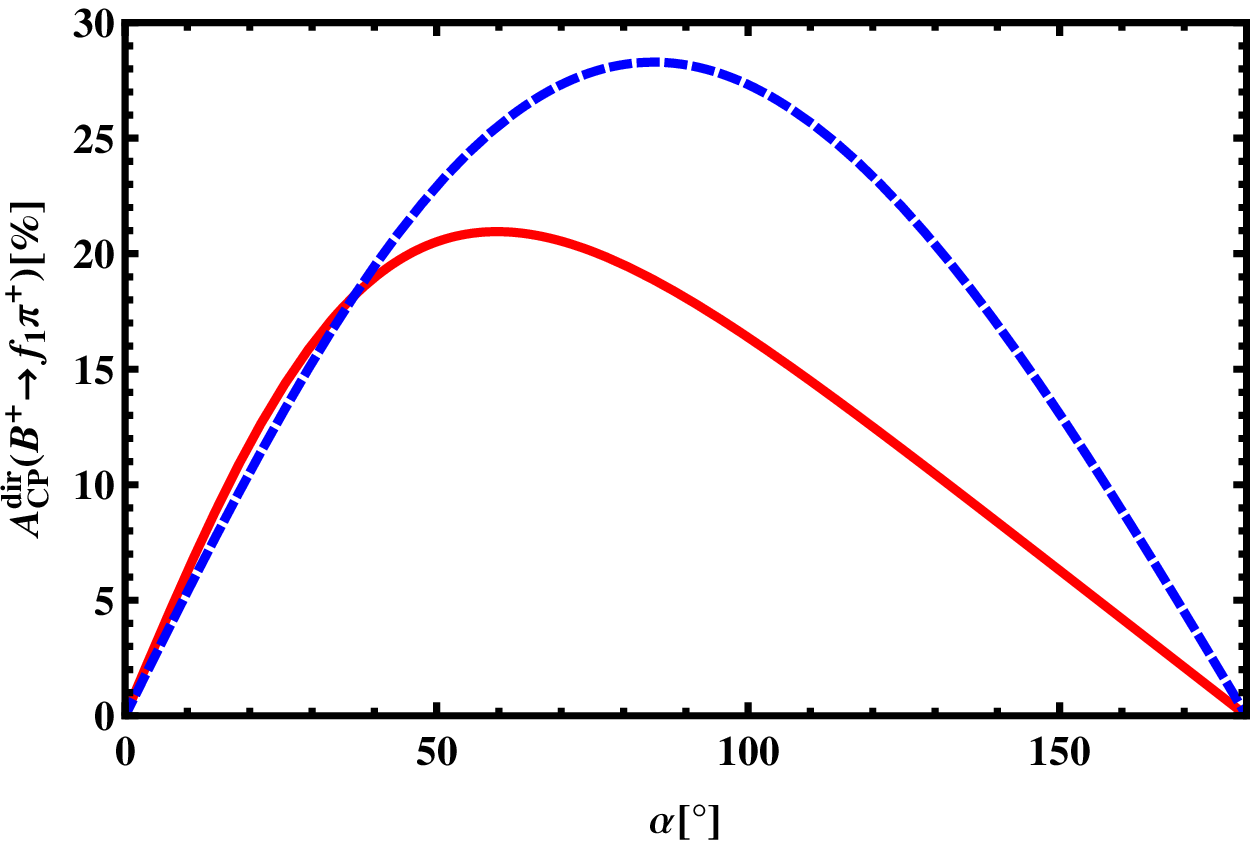}\hspace{0.6cm}
\includegraphics[scale=0.55]{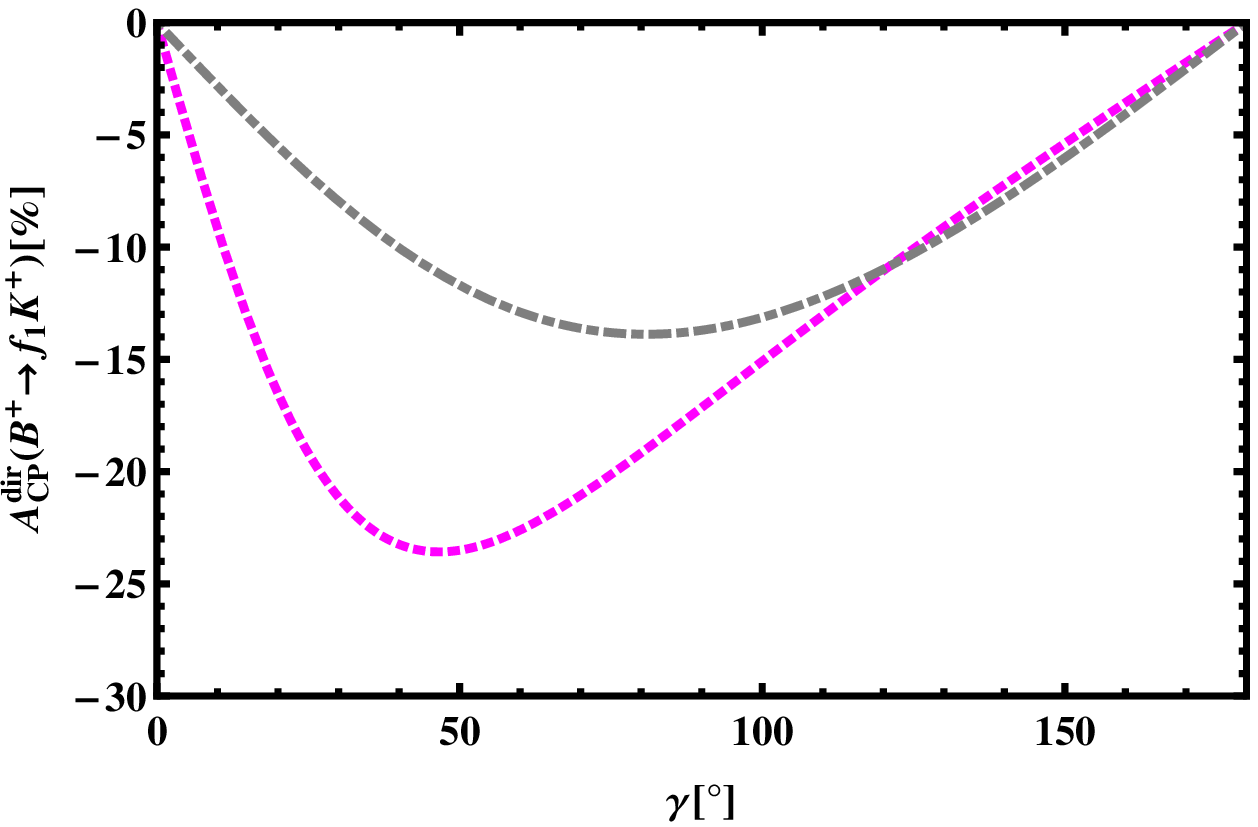}
\caption{(Color online) Dependence on the CKM weak phase $\alpha(\gamma)$ of central values
of the direct \cp-violating asymmetries for $B^+ \to f_1 \pi^+(K^+)$ decays in the pQCD approach.
The red solid\ [blue dashed] line corresponds to the $B^+ \to f_1(1285) \pi^+\ [f_1(1285) \pi^+]$ decay
and the magenta dotted[gray dot-dashed] one corresponds to the $B^+ \to f_1(1285) K^+\ [f_1(1420) K^+])$
 decay, respectively.  }
\label{fig:fig4}
\end{center}
\end{figure}

The central values of the large direct \cp\ violations
for the $B^+ \to f_1(1285) \pi^+$(red solid line) and $B^+ \to f_1(1420) \pi^+$(blue dashed line)
[$B^+ \to f_1(1285) K^+$(magenta dashed line) and $B^+ \to f_1(1420) K^+$(gray dot-dashed line)]
decays as a function of the CKM weak phase $\alpha[\gamma]$ in the pQCD approach have also
been shown in Fig.~\ref{fig:fig4}.
One can find straightforwardly from Fig.~\ref{fig:fig4} that $\acp^{\rm dir}(B^+ \to f_1 \pi^+)$
are large and positive, while $\acp^{\rm dir}(B^+ \to f_1 K^+)$
are large and negative, which are expected to be tested by the experiments in the near future.

It is important to note that the mixing-induced \cp-violating
asymmetries of the $B_d^0 \to f_1 K_S^0$ decays are with the very small uncertainties
as seen clearly in Table~\ref{tab:d-m-cp-bds}, which, as the alternative channels, are expected
to have the supplementary power in reducing the errors of the CKM weak phase $\beta$.
We can write the expression of the \cp-violating parameter $\lambda_{\rm CP}(f_1 K_S^0)$
in an explicit form,
 \beq
\lambda_{\rm CP}(f_1 K_S^0) &=& - \exp{(-2 i \beta)}
\frac{|\lambda_u^s|T_{f_1 K_S^0} \exp{(-i\gamma)} - |\lambda_t^s| P_{f_1 K_S^0}}
{|\lambda_u^s|T_{f_1 K_S^0} \exp{(i\gamma)} - |\lambda_t^s| P_{f_1 K_S^0}}\;.
\label{eq:lambda1}
 \eeq
Here, $|\lambda_u^s| \sim 0.02 \cdot |\lambda_t^s|$ and $T_{f_1 K_S^0}$ is the decay amplitude
arising from the color-suppressed tree diagrams, which will consequently result in
the negligible tree pollution relative to the much larger penguin contributions in the
$B_d^0 \to f_1 K_S^0$ decays, and then $\lambda_{\rm CP}(f_1 K_S^0)
\approx -\exp{(-2i\beta)}$, i.e., $\acp^{\rm mix} = {\cal S}_f \sim \sin{(2\beta_{\rm eff})}$.
In principle, the results should be identical to those measuring
the ${\cal S}_f = -\eta_f \sin2\beta$ from the tree dominated $b \to c \bar c s$
transitions, such as the theoretically cleanest $B_d^0 \to J/\psi K_{S,L}^0$.
However, the $b \to s q \bar q$ decays are potentially contaminated by the
indeed existing tree pollution.  Then the deviation between ${\cal S}_{\rm penguin}$ and
${\cal S}_{\rm c\bar c s}$
can be defined as $\Delta {\cal S} \equiv {\cal S}_{\rm penguin} - {\cal S}_{\rm c\bar c s}$,
which will be helpful to justify the discrepancies as promising new physics signals.
Up to now, the world average value of the ${\cal S}_{\rm c\bar c s}$ at the experimental
aspect is~\cite{Agashe:2014kda}
\beq
\sin2\beta &= & 0.682 \pm 0.019 \;.
\eeq
Then our pQCD predictions of $\sin2\beta_{\rm eff}$ for the $B_d^0 \to f_1 K_S^0$ decays
deviate to the $\sin2\beta$ as
\beq
\Delta{\cal S}_{f_1(1285) K_S^0} & \approx&  0.018^{+0.036}_{-0.035}\;,  \qquad
\Delta{\cal S}_{f_1(1420) K_S^0} \approx 0.017^{+0.031}_{-0.029} \;,
\eeq
which are well below the bound, at most ${\cal O}(0.1)$~\cite{London:1997zk}, 
and can be confronted with stringent tests by the future experiments.

\section{Conclusions and Summary} \label{sec:summary}

In this work, we have studied the \cp-averaged branching ratios and the
\cp-violating asymmetries of 20 charmless hadronic $B \to f_1 P$ decays within
the framework of the pQCD approach. We explicitly evaluated the nonfactorizable
spectator and annihilation types of diagrams, except for the traditional
factorizable emission ones. Based on the quark-flavor mixing of the $f_1(1285)-f_1(1420)$
system with the angle $\phi_{f_1} \sim 24^\circ$ extracted first from the
$B$ meson decays, we calculated the numerical results for the considered
physical observables and made the phenomenological discussions, correspondingly.
The main conclusions of the present paper are as follows:
\begin{enumerate}

\item[(1)]
For the four charged $B^+ \to f_1 \pi^+(\Delta S =0)$ and $f_1 K^+(\Delta S =1)$ decays,
the large \cp-averaged branching ratios $[{\cal O}(10^{-6})]$ together with the large direct \cp\
asymmetries predicted in the pQCD approach are believed to be clearly measurable at the running
LHC and forthcoming Belle II 
experiments in the near future. Furthermore, it is expected
that they could provide supplementary constraints on the CKM weak phase $\alpha\ (\gamma)$ because
of the correspondingly tree-dominant\ (penguin-dominant) contributions to the former(latter) decays. Of course, inferred from the numerical results for the large decay rates theoretically and
the preliminary upper limits for the branching ratios of $B^+ \to f_1 K^+$ modes
experimentally, the region of angle $\phi_{f_1}$ can be deduced as $\phi_{f_1} \in
[20^\circ, 27^\circ]$ by combining with the earlier phenomenological analysis, experimental
measurements and updated LQCD calculations, which provide more evidence for
the dominance of $f_{1q}\ [f_{1s}]$ in $f_1(1285)\ [f_1(1420)]$.

\item[(2)]
Based on the \cp-averaged branching ratios of $B \to f_1 (\pi, K)$ decays calculated in the
pQCD approach, the destructive or constructive interferences between $f_{1q} (\pi, K)$ and $f_{1s} (\pi, K)$ states can be clearly observed and are expected to be confronted with
the future experiments. Also, besides the effects from the $f_{1q}-f_{1s}$ mixing,
the $B_{d/s}^0 \to f_1 (\eta, \eta')$ modes embrace another
set of interferences from $\eta_q-\eta_s$ mixing for $\eta-\eta'$ system simultaneously,
which makes more complicated interactions among the 
four $B_{d(s)}^0 \to f_{1q} \eta_q$, $f_{1q} \eta_s$,
$f_{1s} \eta_q$, and $f_{1s} \eta_s$ states.

\item[(3)]
For the eight neutral $B_{d,s}^0 \to f_1 (\pi^0, \eta, \eta', \bar K^0)$ decays, they are
mediated by the $b \to d$ transitions and dominated by the penguin amplitudes just with
small color-suppressed tree contributions, which then lead to the small \cp-averaged branching
ratios in the order of $10^{-8} \sim 10^{-7}$ that cannot be measured by the experiments
in a short period.

\item[(4)]
The remaining eight neutral $B_{s,d}^0 \to (\pi^0, \eta, \eta', K_S^0)$ modes decay
through $b \to s$ transitions and have large \cp-averaged branching ratios in the order of
$10^{-6} \sim 10^{-5}$, except for $B_s^0 \to f_1 \pi^0$ decays.
The channels with large decay rates are all contributed by the nearly pure penguin amplitudes
with tiny and safely negligible tree pollution, which can be easily accessed at the ongoing
LHCb experiments in the near future.

\item[(5)]
In principle, $B_d^0 \to f_1 K_S^0$ and $B_s^0 \to f_1 (\eta, \eta')$ modes can
serve as the alternative channels to provide more information on the $B_d^0-\bar B_d^0$ and
$B_s^0-\bar B_s^0$ mixing phases from the mixing-induced \cp\ asymmetries ${\cal S}_f$,
respectively. However, the latter
$B_s^0$ decays suffer from large theoretical uncertainties that
 consequently result in the less effective constraints on the mixing phase $\phi_s$.
Fortunately, the former $B_d^0$ ones induced by the $b \to s q \bar q$ decays have
large mixing-induced \cp-violating asymmetries but with very small errors.
The resultant deviations of $\Delta {\cal S}$ for $B_d^0 \to f_1(1285) K_S^0$ and $f_1(1420)
K_S^0$ are around 0.02, which will be stringently examined by the experiments with high
precision.

\item[(6)]
The weak annihilation contributions to these 20 $B \to f_1 P$ decays have been examined
in the pQCD approach. The numerical results show that the sizable effects from annihilation diagrams play important roles in the $B_{u,d} \to f_1(1420) K$, $B_d^0 \to f_1 \eta',
f_1(1420) (\pi, \eta)$, and $B_s^0 \to f_1(1285) \bar K^0, f_1(1420) \eta'$ decays.
The remaining channels do not depend sensitively on the weak annihilation contributions.
The reliability of the evaluations of the weak annihilation diagrams made in the pQCD
approach should be strictly examined by the future experiments, which can help
 to distinguish the different viewpoints on calculating the annihilation
diagrams proposed by the pQCD approach and soft-collinear effective theory, and then
to further understand the annihilation decay mechanism in the heavy $b$-flavored meson
decays.

\item[(7)]
Admittedly, our pQCD results suffer from large theoretical errors induced by the
less constrained hadronic parameters, in particular, from the axial-vector $f_1$
mesons' wave function presently. Meanwhile, only the short-distance contributions
at leading order without considering the final state interactions have been taken
into account. However, the channels such as $B^+ \to f_1 (\pi^+, K^+)$, $B_d^0 \to
f_1 K_S^0$, and $B_s^0 \to f_1 (\eta, \eta')$ with large branching ratios are easily
accessible in the near future measurements with precision at LHCb and/or 
Belle II 
experiments, which are expected in turn to provide useful information on
improving the input quantities; on the other hand, they can help to understand
the mixing angle $\phi_{f_1}$ and the nature of both $f_1$ mesons better and to
identify the reliability of the perturbative evaluations of QCD factorization
and the pQCD approach in these decays involving axial-vector mesons.

\end{enumerate}


\begin{acknowledgments}

X.L. thanks the Institute of Physics at Academia Sinica Taiwan
for the warm hospitality during his visit,
where part of this work was finalized.
This work is supported by the National Natural Science
Foundation of China under Grants No.~11205072, No.~11235005, and No.~11047014,
and by a project funded by the Priority Academic Program Development
of Jiangsu Higher Education Institutions (PAPD),
by the Research Fund of Jiangsu Normal University under Grant No.~11XLR38,
and by the Foundation of Yantai University
under Grant No. WL07052.
\end{acknowledgments}


\begin{appendix}

\section{Hadrons' distribution amplitudes} \label{sec:app-1}

For the $B$ meson, the distribution amplitude in the impact $b$
space has been proposed as~\cite{Keum:2000ph,Lu:2000em}
\beq
\phi_{B}(x,b)&=& N_Bx^2(1-x)^2
\exp\left[-\frac{1}{2}\left(\frac{xm_B}{\omega_b}\right)^2
-\frac{\omega_b^2 b^2}{2}\right] \;,
\eeq
where the normalization factor $N_{B}$
is related to the decay constant $f_{B}$ through Eq.~(\ref{eq:norm}).
The shape parameter $\omega_b$ has been fixed at $\omega_b=0.40$~GeV by using
the rich experimental data on the $B_{u,d}$ mesons with $f_{B_{u,d}}= 0.19$~GeV based on
lots of calculations of form factors~\cite{Lu:2002ny} and other well-known
decay modes of $B_{u,d}$ mesons~\cite{Keum:2000ph,Lu:2000em}
in the pQCD approach in recent years.
Because the $s$ quark is heavier than the $u$ or $d$ quark,
the momentum fraction of the $s$ quark should be a little larger than that
of the $u$ or $d$ quark in the $B_{u,d}$ mesons.
Therefore, by considering a small SU(3) symmetry breaking,
we adopt the shape parameter
$\omega_{bs} = 0.50$~GeV~\cite{Ali:2007ff} with $f_{B_s} = 0.23$~GeV
for the $B_s$ meson, and the
corresponding normalization
constant is $N_{B_s} = 63.67$.
In order to analyze the uncertainties of
theoretical predictions induced
by the inputs, we can vary the shape parameters $\omega_{b}$ and $\omega_{bs}$ by
10\%, i.e., $\omega_b = 0.40 \pm 0.04$~GeV and $\omega_{bs} = 0.50 \pm 0.05$~GeV,
respectively.

The twist-2 pseudoscalar meson distribution amplitude
$\phi_{\pi,K}^A$ and the twist-3 ones $\phi_{\pi,K}^P$ and
$\phi_{\pi,K}^T$ have been parametrized
as~\cite{Chernyak:1983ej,Ball:1998tj,Braun:2004vf}
\beq
\phi_{\pi,K}^A(x) &=& \frac{f_{\pi,K}}{2\sqrt{2N_c}}\, 6x(1-x) \left[1
+ a_1^{\pi,K} C_1^{3/2}(2x-1) +
a_2^{\pi,K}C_2^{3/2}(2x-1)+a_4^{\pi,K}C_4^{3/2}(2x-1)\right] \;,
\\
\phi^P_{\pi,K}(x) &=& \frac{f_{\pi,K}}{2\sqrt{2N_c}}\, \bigg[ 1
+\left(30\eta_3 -\frac{5}{2}\rho_{\pi,K}^2\right) C_2^{1/2}(2x-1) \non &
& \hspace{35mm} -\, 3\left\{ \eta_3\omega_3 +
\frac{9}{20}\rho_{\pi,K}^2(1+6a_2^{\pi,K}) \right\} C_4^{1/2}(2x-1)
\bigg]\;,\\
\phi^T_{\pi,K}(x) &=& \frac{f_{\pi,K}}{2\sqrt{2N_c}}\,
(1-2x)\bigg[ 1 + 6\left(5\eta_3 -\frac{1}{2}\eta_3\omega_3 -
\frac{7}{20}
      \rho_{\pi,K}^2 - \frac{3}{5}\rho_{\pi,K}^2 a_2^{\pi,K} \right)
(1-10x+10x^2) \bigg]\;,\ \ \ \
\eeq
with the Gegenbauer moments
$a_1^{\pi}=0, a_1^{K}=0.17 \pm 0.17, a_2^{\pi,K}= 0.115 \pm 0.115,
a_4^{\pi,K}=-0.015$; the mass ratio
$\rho_{\pi,K}=m_{\pi,K}/m_{0}^{\pi,K}$ and $\rho_{\eta_{q(s)}}=2
m_{q(s)}/m_{qq(ss)}$; and the Gegenbauer polynomials $C_n^{\nu}(t)$,
\begin{eqnarray}
 C_2^{1/2}(t)\,& =&\, \frac{1}{2} \left(3\, t^2-1\right) \;,\;\;\;\;
C_4^{1/2}(t)\, =\, \frac{1}{8} \left(3-30\, t^2+35\, t^4\right) \;,
\nonumber\\
C_1^{3/2}(t)\, &=&\, 3\, t \;, \; \; \;  C_2^{3/2}(t)\, =\,
\frac{3}{2} \left(5\, t^2-1\right) \;, \;\; C_4^{3/2}(t) \,=\,
\frac{15}{8} \left(1-14\, t^2+21\, t^4\right) \;.
\end{eqnarray}
In the above distribution amplitudes for the kaon, the momentum fraction
$x$ is carried by the $s$ quark. For both the pion and kaon, we
choose $\eta_3=0.015$ and $\omega_3=-3$
\cite{Chernyak:1983ej,Ball:1998tj}.

For the axial-vector states $f_{1q(s)}$, its
leading twist light-cone distribution amplitude in the longitudinal polarization
can generally be expanded as the Gegenbauer polynomials~\cite{Yang:2007zt}:
\beq
 \phi_{f_{1q(s)}}(x)&=&\frac{f_{f_{1q(s)}}}{2\sqrt{2N_c}} 6 x  (1-x) \left[ 1
 + a_{2f_{1q(s)}}^\parallel\, \frac{3}{2} ( 5(2x-1)^2  - 1 )
\right]\;.
\eeq

For twist-3 light-cone distribution amplitudes, we use the following
form~\cite{Li:2009tx}:
\beq
\phi_{f_{1q(s)}}^s(x)&=&\frac{f_{f_{1q(s)}}}{4\sqrt{2N_c}} \frac{d}{dx}\Biggl[ 6x
(1-x) (  a_{1f_{1q(s)}}^\perp (2x-1) )\Biggr]\;, \\  
\phi_{f_{1q(s)}}^t(x)&=&\frac{f_{f_{1q(s)}}}{2\sqrt{2N_c}}\Biggl[
\frac{3}{2}\,a_{1f_{1q(s)}}^\perp\,(2x-1) (3 (2x-1)^2-1)\Biggr]\;,
\eeq
where the Gegenbauer
moments are quoted from Ref.~\cite{Yang:2007zt} as 
\beq
f_{1q}\;\; {\rm state:}  \hspace{1.5cm}
a_2^{\parallel}&=&-0.02 \pm 0.02\;,
\qquad
a_1^{\perp}=-1.04 \pm 0.34\;;
\eeq
and
\beq
f_{1s}\;\; {\rm state:}  \hspace{1.5cm}
a_2^{\parallel}&=&-0.04 \pm 0.03\;;
\qquad
a_1^{\perp}= -1.06   \pm 0.36 \;,
\eeq
where the values are taken at $\mu=1$ GeV.

\end{appendix}



\begin{thebibliography}{99}

\bibitem{Feldmann:2014iha}
  T.~Feldmann,
  arXiv:1408.0300 [hep-ph];
  W.~Wang, R.~H.~Li and C.~D.~L\"u,
  Phys.\ Rev.\ D {\bf 78}, 074009 (2008).


\bibitem{Calderon:2007nw}
  G.~Calderon, J.~H.~Munoz and C.~E.~Vera,
  Phys.\ Rev.\ D {\bf 76}, 094019 (2007).

\bibitem{Agashe:2014kda}
  K.~A.~Olive {\it et al.}  [Particle Data Group Collaboration],
  Chin.\ Phys.\ C {\bf 38}, 090001 (2014).

\bibitem{Aaij:2013rja}
  R.~Aaij {\it et al.}  [LHCb Collaboration],
  Phys.\ Rev.\ Lett.\  {\bf 112}, 091802 (2014).

\bibitem{Close:1997nm}
  F.~E.~Close and A.~Kirk,
  Z.\ Phys.\ C {\bf 76}, 469 (1997).


\bibitem{Li:2000dy}
  D.~M.~Li, H.~Yu and Q.~X.~Shen,
  Chin.\ Phys.\ Lett.\  {\bf 17}, 558 (2000).

\bibitem{Liu:2014doa}
  X.~Liu and Z.~J.~Xiao,
  Phys.\ Rev.\ D\ {\bf 89}, 097503 (2014).

\bibitem{Cheng:2011pb}
  H.~Y.~Cheng,
  Phys.\ Lett.\ B {\bf 707}, 116 (2012).


\bibitem{Liu:2010da}
  X.~Liu and Z.~J.~Xiao,
  Phys.\ Rev.\ D {\bf 81}, 074017 (2010);
  Z.~J.~Xiao and X.~Liu,
  Chin.\ Sci.\ Bull.\  {\bf 59}, 3748 (2014).



\bibitem{Keum:2000ph}
  Y.~Y.~Keum, H.-n.~Li and A.~I.~Sanda,
  Phys.\ Lett.\ B {\bf 504}, 6 (2001);
  Phys.\ Rev.\ D {\bf 63}, 054008 (2001).


\bibitem{Lu:2000em}
  C.~D.~L\"u, K.~Ukai and M.~Z.~Yang,
  Phys.\ Rev.\ D {\bf 63}, 074009 (2001).


\bibitem{Li:2003yj}
  H.-n.~Li,
  Prog.\ Part.\ Nucl.\ Phys.\  {\bf 51}, 85 (2003).

\bibitem{Burke:2007zz}
  J.~P.~Burke [BaBar Collaboration],
  J.\ Phys.\ Conf.\ Ser.\  {\bf 110}, 052005 (2008).

\bibitem{Cheng:2007mx}
  H.~Y.~Cheng and K.~C.~Yang,
  Phys.\ Rev.\ D {\bf 76}, 114020 (2007).

\bibitem{Buchalla:1995vs}
  G.~Buchalla, A.J.~Buras and M.E.~Lautenbacher,
  Rev.\ Mod.\ Phys.\  {\bf 68}, 1125 (1996).

\bibitem{Li:2010nn}
  H.~n.~Li, Y.~L.~Shen, Y.~M.~Wang and H.~Zou,
  Phys.\ Rev.\ D {\bf 83}, 054029 (2011);
  H.~n.~Li, Y.~L.~Shen and Y.~M.~Wang,
  Phys.\ Rev.\ D {\bf 85}, 074004 (2012);
  \jhep {\bf 02}, 008 (2013);
  H.~C.~Hu and H.~n.~Li,
  Phys.\ Lett.\ B {\bf 718}, 1351 (2013).

\bibitem{Cheng:2014gba}
  S.~Cheng, Y.~Y.~Fan and Z.~J.~Xiao,
  Phys.\ Rev.\ D {\bf 89}, 054015 (2014);
  S.~Cheng, Y.~Y.~Fan, X.~Yu, C.~D.~L¨¹ and Z.~J.~Xiao,
  Phys.\ Rev.\ D {\bf 89}, 094004 (2014);
  S.~Cheng and Z.~J.~Xiao,
  Phys.\ Rev.\ D {\bf 90}, 076001 (2014);
  arXiv:1409.5947 [hep-ph].


\bibitem{Zhu:2011mm}
  G.~Zhu,
  Phys.\ Lett.\ B {\bf 702}, 408 (2011);
  K.~Wang and G.~Zhu,
  Phys.\ Rev.\ D {\bf 88}, 014043 (2013).


\bibitem{Chang:2014rla}
  Q.~Chang, J.~Sun, Y.~Yang and X.~Li,
  arXiv:1409.1322 [hep-ph];
  Q.~Chang, J.~Sun, Y.~Yang and X.~Li,
  arXiv:1409.2995 [hep-ph].


\bibitem{Chay:2007ep}
  J.~Chay, H.-n.~Li and S.~Mishima,
  Phys.\ Rev.\ D {\bf 78}, 034037 (2008).


\bibitem{Arnesen:2006vb}
  C.M.~Arnesen, Z.~Ligeti, I.Z.~Rothstein and I.W.~Stewart,
  Phys.\ Rev.\ D {\bf 77}, 054006 (2008).


\bibitem{Lu:2002iv}
  C.D.~Lu and K.~Ukai,
  Eur.\ Phys.\ J.\ C {\bf 28}, 305 (2003).


\bibitem{Li:2004ep}
  Y.~Li, C.D.~Lu, Z.J.~Xiao and X.Q.~Yu,
  Phys.\ Rev.\ D {\bf 70}, 034009 (2004).


\bibitem{Ali:2007ff}
  A.~Ali, G.~Kramer, Y.~Li, C.D.~Lu, Y.L.~Shen, W.~Wang and Y.M.~Wang,
  Phys.\ Rev.\ D {\bf 76}, 074018 (2007).


\bibitem{Xiao:2011tx}
  Z.J.~Xiao, W.F.~Wang and Y.Y.~Fan,
  Phys.\ Rev.\ D {\bf 85}, 094003 (2012).



\bibitem{Hong:2005wj}
  B.H.~Hong and C.D.~Lu,
  Sci.\ China G {\bf 49}, 357 (2006);
  H.-n.~Li and S.~Mishima,
  Phys.\ Rev.\ D {\bf 71}, 054025 (2005);
  H.-n.~Li,
  Phys.\ Lett.\ B {\bf 622}, 63 (2005).


\bibitem{Aaltonen:2011jv}
  T.~Aaltonen {\it et al.}  [CDF Collaboration],
  Phys.\ Rev.\ Lett.\  {\bf 108} (2012) 211803.


\bibitem{Ruffini:2013jea}
  F.~Ruffini,
  FERMILAB-THESIS-2013-02.

\bibitem{Aaij:2012as}
  R.~Aaij {\it et al.}  [LHCb Collaboration],
  \jhep {\bf 10}, 037 (2012).


\bibitem{Li:2001ay}
  H.-n.~Li,
  Phys.\ Rev.\ D {\bf 66}, 094010 (2002).


\bibitem{Li:2002mi}
  H.-n.~Li and K.~Ukai,
  Phys.\ Lett.\ B {\bf 555}, 197 (2003).


\bibitem{Botts:1989kf}
  J.~Botts and G.F.~Sterman,
  Nucl.\ Phys.\ B {\bf 325}, 62 (1989).


\bibitem{Li:1992nu}
  H.-n.~Li and G.F.~Sterman,
  Nucl.\ Phys.\ B {\bf 381}, 129 (1992).



\bibitem{Lu:2002ny}
  C.D.~Lu and M.Z.~Yang,
  Eur.\ Phys.\ J.\ C {\bf 28}, 515 (2003).


\bibitem{Chernyak:1983ej}
  V.~L.~Chernyak and A.~R.~Zhitnitsky,
  Phys.\ Rept.\  {\bf 112}, 173 (1984);
A.~R.~Zhitnitsky, I.~R.~Zhitnitsky and V.~L.~Chernyak,
  Sov.\ J.\ Nucl.\ Phys.\  {\bf 41}, 284 (1985)
  [Yad.\ Fiz.\  {\bf 41}, 445 (1985)];
  V.~M.~Braun and I.~E.~Filyanov,
  Z.\ Phys.\ C {\bf 44}, 157 (1989)
  [Sov.\ J.\ Nucl.\ Phys.\  {\bf 50}, 511 (1989)]
  [Yad.\ Fiz.\  {\bf 50}, 818 (1989)];
  V.~M.~Braun and I.~E.~Filyanov,
  Z.\ Phys.\ C {\bf 48}, 239 (1990)
  [Sov.\ J.\ Nucl.\ Phys.\  {\bf 52}, 126 (1990)]
  [Yad.\ Fiz.\  {\bf 52}, 199 (1990)].


\bibitem{Ball:1998tj}
  P.~Ball,
  \jhep {\bf 09}, 005 (1998);
  P.~Ball,
  \jhep {\bf 01}, 010 (1999).


\bibitem{Yang:2007zt}
  K.C.~Yang,
  Nucl.\ Phys.\ B {\bf 776}, 187 (2007).


\bibitem{Li:2009tx}
  R.H.~Li, C.D.~Lu and W.~Wang,
  Phys.\ Rev.\ D {\bf 79}, 034014 (2009).



\bibitem{Liu:2005mm}
  X.~Liu, H.~s.~Wang, Z.~j.~Xiao, L.~Guo and C.~D.~Lu,
  Phys.\ Rev.\ D {\bf 73}, 074002 (2006).



\bibitem{Feldmann:1998vh}
  T.~Feldmann, P.~Kroll and B.~Stech,
  Phys.\ Rev.\ D {\bf 58}, 114006 (1998).

\bibitem{Escribano:2005qq}
  R.~Escribano and J.~M.~Frere,
  \jhep {\bf 06}, 029 (2005);
  J.~Schechter, A.~Subbaraman and H.~Weigel,
  Phys.\ Rev.\ D {\bf 48}, 339 (1993)

\bibitem{DiDonato:2011kr}
  C.~Di Donato, G.~Ricciardi and I.~Bigi,
  Phys.\ Rev.\ D {\bf 85}, 013016 (2012).




\bibitem{Wang:2005bk}
H.~s.~Wang, X.~Liu, Z.~j.~Xiao, L.~b.~Guo and C.~D.~Lu,
  Nucl.\ Phys.\ B {\bf 738}, 243 (2006);
  Z.~j.~Xiao, D.~q.~Guo and X.~f.~Chen,
  Phys.\ Rev.\ D {\bf 75}, 014018 (2007);
  Z.~j.~Xiao, X.~Liu and H.~s.~Wang,
  Phys.\ Rev.\ D {\bf 75}, 034017 (2007);
  D.~Q.~Guo, X.~F.~Chen and Z.~J.~Xiao,
  Phys.\ Rev.\ D {\bf 75}, 054033 (2007);
  Z.~J.~Xiao, Z.~Q.~Zhang, X.~Liu and L.~B.~Guo,
  Phys.\ Rev.\ D {\bf 78}, 114001 (2008);
  X.~Liu, H.~n.~Li and Z.~J.~Xiao,
  Phys.\ Rev.\ D {\bf 86}, 011501 (2012);
  Y.~Y.~Fan, W.~F.~Wang, S.~Cheng and Z.~J.~Xiao,
  Phys.\ Rev.\ D {\bf 87}, 094003 (2013).


\bibitem{Charng:2006zj}
  Y.~Y.~Charng, T.~Kurimoto and H.~n.~Li,
  Phys.\ Rev.\ D {\bf 74}, 074024 (2006)
  [Erratum-ibid.\ D {\bf 78}, 059901 (2008)].

\bibitem{Escribano:2007cd}
  R.~Escribano and J.~Nadal,
  \jhep {\bf 05}, 006 (2007).


\bibitem{Yang:2010ah}
  K.~C.~Yang,
  Phys.\ Rev.\ D {\bf 84}, 034035 (2011).


\bibitem{Cheng:2013cwa}
  H.~Y.~Cheng,
  {\it Proc. Sci.}, Hadron2013, 090 (2014).

\bibitem{Verma:2011yw}
  R.~C.~Verma,
  J.\ Phys.\ G {\bf 39}, 025005 (2012).

\bibitem{Gidal:1987bn}
  G.~Gidal, J.~Boyer, F.~Butler, D.~Cords, G.~S.~Abrams, D.~Amidei, A.~R.~Baden and T.~Barklow {\it et al.},
  Phys.\ Rev.\ Lett.\  {\bf 59}, 2012 (1987).

\bibitem{Dudek:2013yja}
  J.~J.~Dudek, R.~G.~Edwards, P.~Guo and C.~E.~Thomas,
  Phys.\  Rev.\  D  {\bf 88}, 094505 (2013)


\bibitem{Beringer:1900zz}
  J.~Beringer {\it et al.}  [Particle Data Group Collaboration],
  Phys.\ Rev.\ D {\bf 86}, 010001 (2012).


\bibitem{Donoghue:1986wv}
  J.~F.~Donoghue, B.~R.~Holstein and Y.~C.~R.~Lin,
  Phys.\ Rev.\ Lett.\  {\bf 55}, 2766 (1985)
  [Erratum-ibid.\  {\bf 61}, 1527 (1988)].


\bibitem{Beneke:1998sy}
  M.~Beneke, G.~Buchalla, C.~Greub, A.~Lenz and U.~Nierste,
  Phys.\ Lett.\ B {\bf 459}, 631 (1999).


\bibitem{Fernandez:2006qx}
  L.~Fernandez,
  CERN-THESIS-2006-042.

\bibitem{London:1997zk}
  D.~London and A.~Soni,
  Phys.\ Lett.\ B {\bf 407}, 61 (1997).


\bibitem{Braun:2004vf}
  V.~M.~Braun and A.~Lenz,
  Phys.\ Rev.\ D {\bf 70}, 074020 (2004);
  P.~Ball and A.~N.~Talbot,
  \jhep {\bf 06}, 063 (2005);
  P.~Ball and R.~Zwicky,
  Phys.\ Lett.\ B {\bf 633}, 289 (2006);
  A.~Khodjamirian, T.~Mannel and M.~Melcher,
  Phys.\ Rev.\ D {\bf 70}, 094002 (2004).


\end{thebibliography}
\end{document}